\def\equationautorefname~#1\null{Equation (#1)\null}
\def\sectionautorefname~#1\null{Section #1\null}
\def\subsectionautorefname~#1\null{Section #1\null}
\def\subsubsectionautorefname~#1\null{Section #1\null}
\def\figureautorefname~#1\null{Figure #1\null}

\documentclass[twocolumn]{aastex63}
\usepackage{amsmath}
\usepackage{xparse}
\usepackage{hyperref}
\usepackage[flushleft]{threeparttable}
\usepackage{bm}
\usepackage{xcolor}
\usepackage{ulem}%

\newcommand*{\ar}{\autoref}
\newcommand*{\ct}{\citet}
\newcommand*{\ctp}{\citep}
\newcommand*{\cta}{\citealp}

\received{}
\revised{}
\accepted{\today}
\graphicspath{{./}{figures/}}

\makeatletter
\@ifpackageloaded{caption}{}{%
  \providecommand{\captionsetup}[1]{}
}
\makeatother

\begin{document}

\title{Ultraviolet Signatures of Jet–Ejecta Interaction in Early Kilonovae: Prediction from Realistic Atomic Opacities} 

\correspondingauthor{Smaranika Banerjee}
\email{smaranika.banerjee@astro.su.se}

\author[0000-0001-6595-2238]{Smaranika Banerjee}
\affiliation{Institut f\"ur Kernphysik, Technische Universit\"at Darmstadt, Schlossgartenstr.~2, 64289 Darmstadt, Germany
}

\author[0000-0003-2866-4522]{Hamid Hamidani}
\affiliation{Astronomical Institute, Tohoku University, Aoba, Sendai 980-8578, Japan}

\author[0000-0003-4443-6984]{Kyohei Kawaguchi}
\affiliation{Max Planck Institute for Gravitational Physics (Albert Einstein Institute), Am M\"{u}hlenberg 1,
Potsdam-Golm, 14476, Germany;Center of Gravitational Physics and Quantum Information,
 Yukawa Institute for Theoretical Physics, 
Kyoto University, Kyoto, 606-8502, Japan}

\author[0000-0001-8253-6850]{Masaomi Tanaka}
\affiliation{Astronomical Institute, Tohoku University, Aoba, Sendai 980-8578, Japan}
\affiliation{Division for the Establishment of Frontier Sciences, Organization for Advanced Studies,
Tohoku University, Sendai 980-8577, Japan}

\begin{abstract}
We investigate the signature of the jet-ejecta interaction in early kilonova ($t\le1$ day) using detailed atomic opacities developed in \ct{Smaranikab20, Smaranikab24}, appropriate for early times ($t \sim 1$ hour after the merger). We explore jets with different powers and opening angles.
We find that the presence of the jet shifts the spectral peak to longer wavelengths, with the strongest effect near the polar viewing angle. This occurs because the jet creates a thin, low-density outer layer ahead of the bulk ejecta. The opacity of this layer can be as high as $\kappa \sim 200\,\rm cm^2\,g^{-1}$, causing photons to escape from cooler, faster-moving outer layer rather than from the hot inner ejecta. The bolometric light curves likewise exhibit a clear imprint of the jet–ejecta interaction, showing suppressed early-time luminosity near polar viewing angles compared to the equatorial one, as the photosphere resides in this thin layer where radioactive heating is lower than in the bulk ejecta. 
These signatures are also evident in multi-color light curves, particularly in the ultraviolet and $u$-bands. For example, in the \textit{Swift} UVW2 band at $t \simeq 0.15$ days for a source at 100 Mpc, the ultraviolet luminosities can reach $\sim$ 19.5 mag if no jet is present, while the presence of the jet can make it fainter by $\sim 2.5$ mag. The strongest observational signature occurs in the \textit{UVEX} NUV, \textit{Swift} UVW2, and UVM2 bands, which remains detectable out to viewing angles of $\sim 60^\circ$ for $t \lesssim 1$ days. Rapid follow-up with future ultraviolet facilities, such as \textit{ULTRASAT} and \textit{UVEX}, will provide powerful probes of jet–ejecta interaction through early-time kilonova observations.

\end{abstract}


\section{Introduction} \label{sec:intro}
\begin{table*}[!htbp]
\centering
\caption{Summary of the jet-interacted ejecta models adopted from \cite{Hamidani23a,Hamidani23b}. 
In all cases, the jet is assumed to remain active for 2~s. 
The bulk ejecta into which the jet is launched follows a power-law density profile with index $-2$. 
The table lists, from left to right: the model names (defined by the jet power and opening angle);
the isotropic-equivalent luminosity of the jet engine, $L_{\rm iso} = \dfrac{2L_j}{1-\cos\theta_{j}} \simeq 4L_j/\theta_{j}^2$, 
where $L_{j}$ is the true (one-sided) jet power; the initial jet opening angle $\theta_{j}$; 
the outcome of whether the jet hydrodynamically breaks out of the ejecta;
and the mass of the thin outer layer, defined as the ejecta/cocoon mass with velocities larger than the edge of the bulk ejecta. 
}
\begin{tabular}{lcccc}
\hline
Model name & $L_{\rm iso}$ [erg s$^{-1}$] & $\theta_{j}$ [$^{\circ}$] & Jet successfully broke out? & $M(>v_{\rm ej, max})$ [$M_\odot$]\\
\hline
PL     & —                & —    & —   & — \\
HE-NJ  & $5\times10^{50}$ & 6.8  & Yes & $2.1\times10^{-5}$\\
LE-NJ  & $1\times10^{50}$ & 6.8  & Yes & $2.0\times10^{-5}$\\
HE-WJ  & $5\times10^{50}$ & 18.0 & Yes & $1.1\times10^{-4}$\\
LE-WJ  & $1\times10^{50}$ & 18.0 & No  & $5.9\times10^{-5}$\\
\hline
\end{tabular}
\label{tab:model}
\end{table*}

On August 17, 2017, the first binary neutron star merger was detected via the multimessenger signal,
simultaneously in gravitational waves (GW170817, \citealt{Abbott17a}) and electromagnetic counterparts across the spectrum. The electromagnetic signal includes the short ($T_{90}\le 2$ s) gamma-ray burst (GRB170817A, e.g., \cta{Connaughton17, Savchenko17, Tanvir17}) and the kilonova, the transient powered by the freshly synthesized heavy elements (e.g., \cta{Li98, Metzger10}), AT2017gfo (e.g., \citealt{Coulter17, Yang17, Valenti17, Cowperthwaite17, Smartt17, Drout17, Utsumi17}).

This observation has marked the watershed moment in many aspects; for example,
the detection of the kilonova confirms that these are one of the sites of heavy elements synthesis,
proving theoretical predictions (e.g., \cta{Lattimer74, Eichler89, Freiburghaus99, Korobkin12, Wanajo14}). Moreover, the discovery also validates the long-standing hypothesis regarding these mergers to be connected with short GRBs (e.g., \cta {Eichler89, Paczynski91}).
In the following years, additional kilonova candidates have been reported in association with both short and long-duration ($T_{90} > 2$~s) GRBs, including GRB~211211A (e.g., \cta{Rastinejad22}) and GRB~230307A (e.g., \cta{Levan24}). These events further support the picture in which relativistic jets are launched following compact object mergers.

Although the association between compact object mergers and relativistic jet launching is now well established, the precise timing of jet formation remains uncertain. For instance, jet launching may be delayed if the magnetic field around the central compact object requires time to amplify to the strength necessary to drive a relativistic outflow (e.g., \cta{Lazzati20}). In such scenarios, the jet propagates through merger ejecta that have already been launched and have started expanding. As the jet drills through the dynamical ejecta, it reshapes the density and angular structure of the outflow (e.g., \cta{Duffell18, Hamidani20, Nativi21, Hamidani23a, Hamidani23b}). 

Previous studies have shown that these jet-induced modifications to the ejecta can leave observable imprints on kilonova emission \ctp{Klion21, Nativi21, Shrestha23}. However, these studies are limited by their use of approximate treatments of the microphysics. During the photospheric phase, the emergent light curves and spectra are primarily governed by the wavelength- and time-dependent opacity of the ejecta, which is determined by the atomic properties of the freshly synthesized $r$-process elements (e.g., \cta{Metzger10, Kasen13, Tanaka13}). Consequently, reliable predictions of kilonova emission demand accurate atomic opacities that consistently capture the evolving thermodynamic and ionization conditions of the expanding ejecta.

At early times, the ejecta are dense and have high temperature. For a GW170817-like kilonova, temperatures can reach $T \sim 10^{5}$~K at $t \sim 1$~hour after merger \ctp{Smaranikab20, Smaranikab22}, leading to highly ionized material, with charge states reaching up to XI \ctp{Smaranikab20}. Modeling the early-time emission, therefore, requires time- and wavelength-dependent opacities for highly ionized heavy elements. However, previous studies of jet-ejecta interaction effects on kilonova light curves have relied on simplified opacity treatments. For example, \citet{Klion21} employed a gray-opacity approximation, while \citet{Nativi21} adopted analytic prescriptions based on low-ionization ($\mathrm{I}$-$\mathrm{IV}$) opacities of a limited set of $r$-process elements \ctp{Tanaka18}. More recently, \citet{Shrestha23} used detailed opacities for elements from Ca - Ra ($Z=20-88$, \cta{Tanaka20a}), but restricted to ionization states up to IV. As demonstrated by \ct{Smaranikab20, Smaranikab22, Smaranikab24}, such treatments are insufficient for the earliest epochs ($t \lesssim 1$~day), when higher ionization stages dominate the opacity.

In this paper, we present kilonova light curves from jet-interacted ejecta in binary neutron star mergers, employing the atomic opacities of \citet{Smaranikab20, Smaranikab24} that are appropriate for the early phase ($t \leq 1$ day). For the jet-interacted ejecta models, we adopt the calculations from \citet{Hamidani23a, Hamidani23b}. The structure of the paper is as follows: in \autoref{sec:ej_struc}, we describe the ejecta structure after modification by the jet; in \autoref{sec:rad_tr}, we provide details of the radiative-transfer simulations; in \autoref{sec:res}, we present the results; in \autoref{sec:disc}, we discuss the detectability of kilonova light curves relative to the afterglow and compare our findings with previous works; and finally in \autoref{sec:conc}, we summarize our conclusions. All magnitudes in this paper are given in the AB system.

\section{Ejecta structure} \label{sec:ej_struc}

\begin{figure*}[!htbp]
  \centering
  \setlength{\tabcolsep}{2pt}%
  \renewcommand{\arraystretch}{0}%
  \captionsetup{font=small, width=\textwidth, belowskip=0pt, aboveskip=4pt}
\begin{tabular}{@{}cc@{}}
    \includegraphics[width=0.45\textwidth]{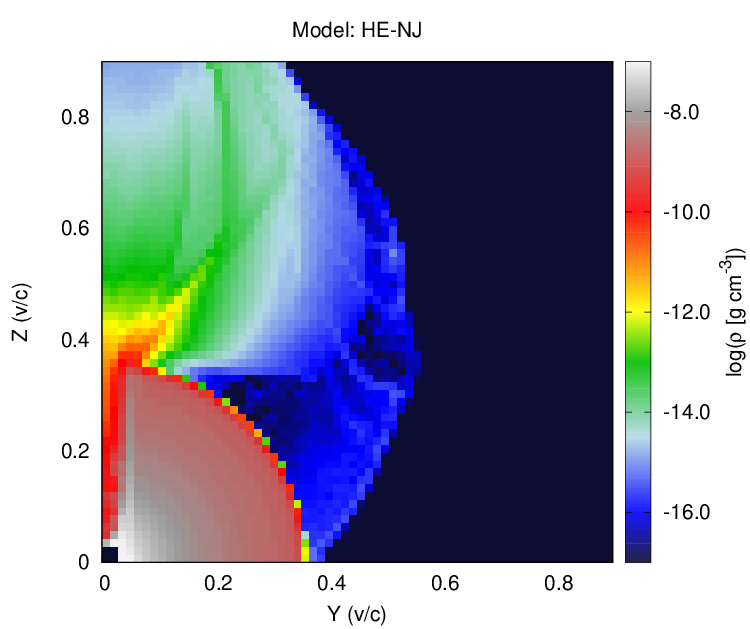} &    
    \includegraphics[width=0.45\textwidth]{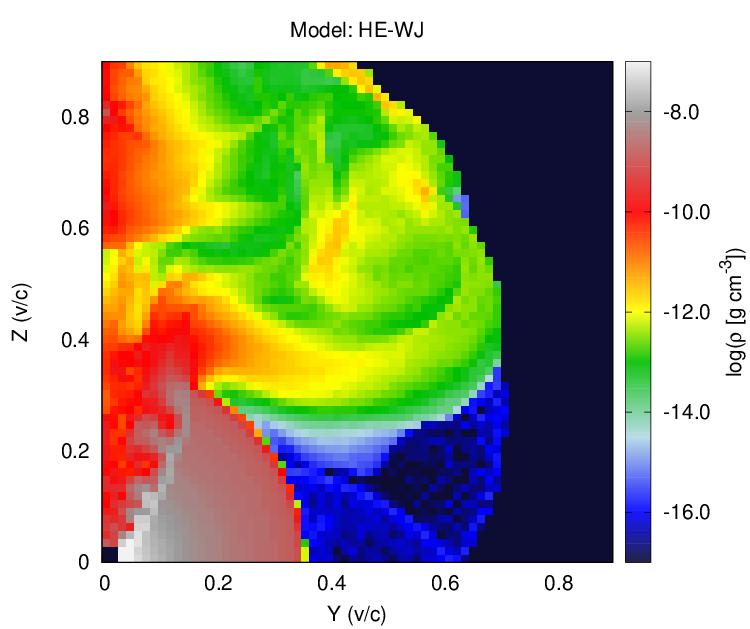}
  \end{tabular}
    \caption{Two-dimensional ejecta structures for different models at $t = 0.03$ days, which is the starting point of our radiative-transfer simulations. The hydrodynamic simulation results are post-processed by mapping them onto an axisymmetric velocity grid with the resolution adopted for the radiative-transfer calculations. Moreover, we exclude all material with $v > 0.9c$. \textbf{Left:} ejecta after interaction with the narrow jet (NJ); \textbf{Right:} ejecta after interaction with the wide jet (WJ). Only the high-luminosity jet models (HE) are shown here. The jet–ejecta interaction produces a thin layer at the leading edge of the ejecta, particularly along the polar direction, while the equatorial regions, being far from the jet core, remain largely unaffected. The black central region denotes material with $v < 0.025c$. This component is omitted from our simulations, as its low mass and extreme optical depth make its contribution to the early-time light curves negligible.}
\label{fig:den1}
\end{figure*}

To model the kilonova emission from jet-modified ejecta in neutron star mergers, we use the two-dimensional hydrodynamical simulations that follow the propagation of relativistic jets through merger ejecta \ctp{Hamidani23a, Hamidani23b}. These simulations provide a physically motivated description of the jet–ejecta interaction and allow us to reconstruct the resulting outflow geometry and composition.
The initial conditions, post-processing procedures, and the ejecta structures adopted for our radiative-transfer calculations are described in the following subsections. A more detailed account of the computational methods underlying these simulations can be found in \cite{Hamidani23a} and \ct{Hamidani23b}.
\subsection{Jet initial conditions}
In hydrodynamical simulations, the jet is assumed to be launched as a consequence of accretion onto the newly formed central compact object. The exact jet launch time depends on the formation timescale, nature, and properties of the remnant, all of which remain uncertain \citep[e.g.,][]{Kiuchi19, Kiuchi23}. For simplicity, we adopt a fiducial launch time of $\simeq 0.16$~s after the merger, consistent with the setup in \citet{Hamidani23a, Hamidani23b}. This choice is also compatible with observational constraints from GRB170817A, which require $t_{\rm launch} \leq 1.3$ s once the jet breakout time is taken into account \citep{Hamidani20}. We further assume that the central engine remains active for $\sim 2$~s, consistent with both the typical duration of short GRBs (e.g., \cta{Kouveliotou93}) and the observed duration of GRB170817A \citep[e.g.,][]{Goldstein17}. 

We investigate a set of jet models by varying both jet power and opening angle. The jet initial opening angle is considered to be $\theta_{j} = 6.8^\circ$ and $\theta_{j} = 18^\circ$ (NJ and WJ models), motivated by the range of opening angles typically inferred for GRBs (e.g., \cta{Fong15, Escorial23}). Similarly, the isotropic-equivalent luminosity of the jet is assumed to be $L_{\rm iso} = 10^{50}$ and $5\times10^{50}\,\mathrm{erg\,s^{-1}}$ (LE and HE models), consistent with the characteristic energy range observed in GRBs \citep{Fong15}. A summary of the adopted models is provided in \ar{tab:model}.
\begin{figure*}[!htbp]
  \centering
  \setlength{\tabcolsep}{2pt}%
  \renewcommand{\arraystretch}{0}%
  \captionsetup{font=small, width=\textwidth, belowskip=0pt, aboveskip=4pt}
\begin{tabular}{@{}cc@{}}
    \includegraphics[width=0.45\textwidth]{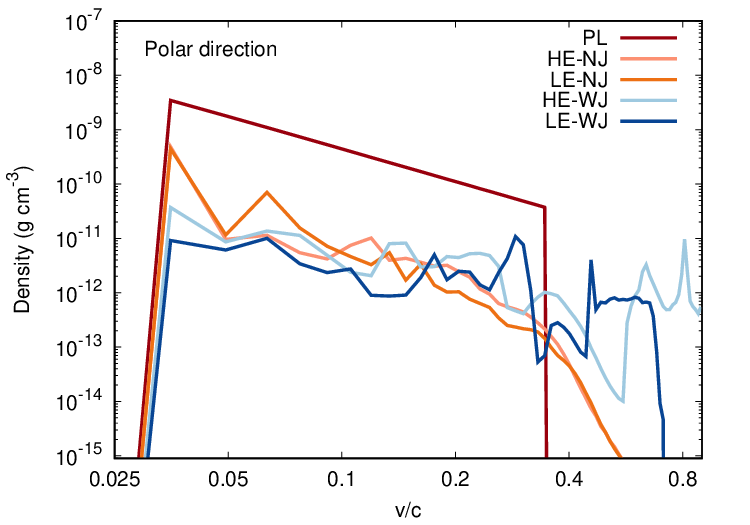} &    
    \includegraphics[width=0.45\textwidth]{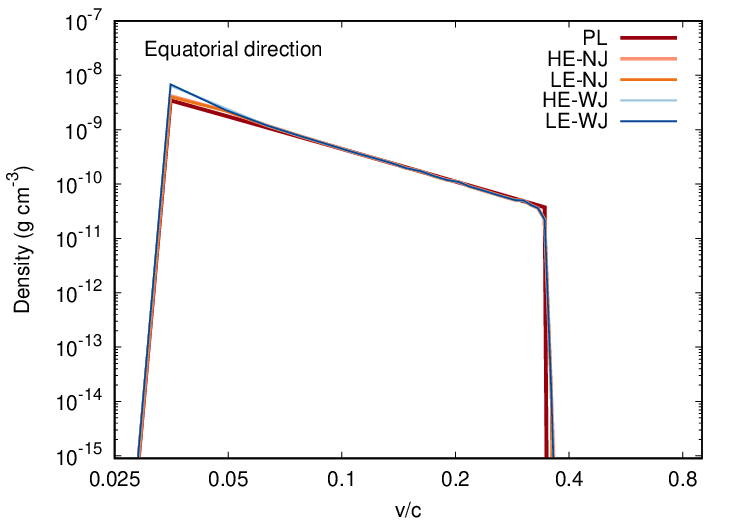}
  \end{tabular}
  
  \caption{One-dimensional ejecta structures for all models shown in \ar{tab:model} at $t = 0.03$ days, marking the starting point of our radiative-transfer simulations, shown along the polar ($\theta_{v} = 0-14^{\circ}$, left) and equatorial ($\theta_{v} = 88-90^{\circ}$, right) directions. Different colors indicate different models, including the PL model without jet interaction. Both the narrow- and wide-jet cases (NJ and WJ) exhibit significant modifications to the density structure relative to the PL model, most prominently along the polar direction. In both cases, a thin layer develops at the leading edge of the polar ejecta, while the equatorial ejecta remains largely unchanged. For the narrow-jet models (HE-NJ and LE-NJ), this layer follows a steeply declining power-law–like profile, whereas in the wide-jet models, the density distribution differs owing to the broader interaction induced by the larger opening angle.}
  \label{fig:den2}
\end{figure*}

\subsection{Initial ejecta structure}
After its launch, the jet propagates through the merger ejecta, modifying its structure along the way. 
This interaction is most pronounced in the polar regions, where the ejecta is strongly affected. 
For the jet-launch timescale adopted in this work, i.e., $t_{\rm jet} = 0.16$~s after the merger,
these regions are assumed to be dominated by post-merger ejecta \citep[e.g.,][]{Fujibayashi18, Fujibayashi20b}. 
Motivated by this picture, we model the ejecta as a steady wind-driven outflow from the accretion disk, resulting in a power-law density profile,  
\begin{equation}
    \rho(r) = \rho_0 \left(\frac{r}{r_0}\right)^{-2},
\end{equation}
where $r_0 \sim 1.2\times 10^8$~cm is the inner boundary of the hydrodynamical simulations. 
The density at this boundary is given by 
\[
\rho_0 = \frac{M_{\rm ej}}{4\pi r_0^2(r_{\rm max}-r_0)}.
\] 
Here $r_{\rm max} = v_{\rm max} t_0$ and $t_0 \simeq 0.16$~s. The ejecta velocity range is taken to be $v_{\rm ej} = 0.025c$ to $0.35c$. The upper limit is motivated by root mean squared velocities from previous simulations \citep[e.g.,][]{Hotokezaka13}.

We adopt the total ejecta mass to be $M_{\rm ej} = 0.01 M_{\odot}$. This lies on the lower end of typical disk wind ejecta mass.
However, the mass adopted here is good enough to represent the earlier jet-propagated ejecta environment, when the majority of the emission appears from the outermost layer. Hence, we consider such ejecta structure, since the early-time ($t\le1$ day) kilonova is the major focus of this work.

\subsection{Post-processing the ejecta}
In the hydrodynamic simulations adopted from \ct{Hamidani23a, Hamidani23b}, the evolution of the jet and ejecta is followed up to $t - t_{0} \sim 10$~s after the merger, as the hydrodynamic calculations are computationally expensive. Within this time span, the ejecta have not yet reached homologous expansion, in which the velocity field satisfies $v = r/t$. 
Since homologous expansion is a key assumption in our radiative-transfer calculations, we therefore post-process the ejecta to extrapolate its structure to later times, when the homologous phase is established. 

For the post-processing of the ejecta, we start from the final snapshot of the hydrodynamic simulations ($t - t_{0} \sim 10$~s).
From this snapshot, we extract the fluid properties in each grid cell, including the rest-mass density, velocity, and specific enthalpy ($h$). 
We then extrapolate these quantities to later times, assuming that sufficiently after jet breakout the fluid elements evolve ballistically,
without further interactions, and undergo only adiabatic expansion.
In this regime, the velocity of each fluid element asymptotically approaches terminal velocity, $v_\infty \simeq c\sqrt{1 - \Gamma_\infty^{-2}}$,
where the terminal Lorentz factor is given by $\Gamma_\infty \simeq h\Gamma$, following Bernoulli’s law. 
We evolve the system to $t \sim 10^{3}$~s after the merger, by which time the ejecta have reached homologous expansion. 
Finally, the post-processed ejecta are mapped from the high-resolution hydrodynamical grid ($512 \times 9200$) onto the radiative-transfer grid ($64\times64$) to obtain the ejecta structure used in our simulations.

\subsection{Final ejecta structure}
\ar{fig:den1} shows the two-dimensional density structure of the ejecta at $t = 0.03$ days, the starting point of our radiative-transfer simulations. The left and right panels correspond to models in which the ejecta interact with a narrow and a wide jet (NJ and WJ), respectively. For simplicity, only the ejecta interacted with the high-luminosity jets ($L_{\rm iso} = 5 \times 10^{50}$ erg s$^{-1}$; HE models) are shown. We impose a maximum ejecta velocity limit of $v_{\rm ej} = 0.9c$, which does not affect the radiative-transfer results since the mass beyond this limit is negligible ($M_{v_{\rm ej}>0.9c} \lesssim 10^{-6} M_{\odot}$; \citealt{Hamidani23a, Hamidani23b}).

In both narrow and wide jet cases (HE-NJ, HE-WJ), the jet–ejecta interaction significantly alters the structure, producing a thin layer ahead of the bulk ejecta, most prominently along the polar direction. By contrast, modifications in the equatorial direction are minimal, consistent with its large angular separation from the jet core. 

A high-density polar plug is visible in the density maps, most prominently in the wide-jet case at velocities $v_{\rm ej} > 0.6c$. This feature arises from the 2D axisymmetric approximation in the hydrodynamic simulations and may affect the jet dynamics. Considering that this plug consists of fast-moving outflow ($v_{\rm ej} > 0.6c$) confined to the vicinity of the jet axis ($\lesssim10^{\circ}$), its presence (i.e., high density) may influence the very early-time KN emission ($\lesssim 0.1$ day or $\lesssim 2$ hours) for observers with a line of sight aligned with the jet axis. However, its impact on the later epochs ($t\gtrsim 0.1$ days) should be non-significant.

To compare the ejecta structure across models, we show the one-dimensional ejecta profiles for all models, including the power-law (PL) model without jet interaction (\ar{tab:model}), at $t = 0.03$ days in \autoref{fig:den2}. The figure displays the density distributions along both the polar (left) and equatorial (right) directions.

The jet-interacted ejecta models exhibit significantly modified density profiles, especially along the polar direction. At velocities below the fiducial maximum of the wind ejecta ($v_{\rm ej} < 0.35c$), the densities are reduced, with the displaced inner material redistributing into a thin outer layer up to a higher velocity, with the exact value depending on the model. In the narrow-jet cases (NJ), the density of this layer declines steeply with velocity, while in the wide-jet cases (WJ), the density structure becomes more complex owing to the interaction happening in the broader region of the ejecta. Along the equatorial direction, the density profiles remain largely unchanged, with only minor variations in the innermost regions across different models, and their peak values remain close to those of the PL model. Note that we see a sharp drop in the inner and outer boundary due to the initial density chosen as the power-law structure (Equation 1).

\section{Radiative transfer simulations} \label{sec:rad_tr}
Using the density structures described in \ar{sec:ej_struc}, we model the kilonova emission from $t = 0.03$ days with a time- and wavelength-dependent Monte Carlo radiative-transfer code \ctp{Tanaka13, Tanaka18, Kawaguchi18, Kawaguchi21}. The code computes the light curves and spectra, given the density distribution, elemental abundances, and radioactive heating rate.

\subsection{Abundances} \label{sec:abun}

In a neutron star merger, masses are ejected in several different channels \ctp{Shibata19}, producing multiple ejecta components, each of which has a different electron fraction ($Y_{\rm{e}}$), i.e., the electron to baryon ratio in the ejecta. For example, tidal dynamical ejecta, which is the mass ejected within the dynamical timescale due to the tidal disruption, has relatively lower electron fraction $Y_{\rm{e}}<0.25$, whereas the polar dynamical ejecta, which is the ejecta emitted towards the polar direction due to the shock between the interface of the neutron stars have relatively higher electron fraction $Y_{\rm{e}}>0.25$ (e.g., \cta{Bauswein13, Just15, Sekiguchi15, Just22}). In addition, mass outflows from the accretion disk formed around the central remnant, which are ejected more isotropically, exhibit a broad range of electron fractions, $Y_{\rm e}$ (e.g., \cta{Fernandez13, Metzger14, Miller19, Fernandez14, Perego14, Lippuner17, Fujibayashi18, Fujibayashi20a, Fujibayashi20b, Fujibayashi20c}).

In this work, we consider mainly the disk wind part of the ejecta, through which the jet moves after getting launched at $t-t_{0} =0.16$ s (see \ar{sec:ej_struc}). We adopt the electron fraction to be $Y_{\rm{e}} = 0.30 - 0.40$, consistent with the possible $Y_{\rm{e}}$ range in the disk wind ejecta.
We take elemental abundances and heating rates from the $r$-process nucleosynthesis results of \citet{Wanajo18}. A flat mass distribution is assumed across the range of $Y_e$ values. For the sake of simplicity, we assume the $Y_{\rm{e}}$ to be the same and the abundances are homogeneous throughout the ejecta.

\subsection{Heating} \label{sec:rad_heat}
The ejecta from a compact-object merger can be heated through multiple channels, including (i) the radioactive decay of freshly synthesized $r$-process nuclei,
(ii) shock or cocoon heating resulting from jet–ejecta interaction, and (iii) energy injection from sustained central-engine activity \citep[see][]{Metzger20}.
In this work, we neglect any long-term central-engine contribution and compare only the first two mechanisms as potential heating sources.

The thermal energy input from radioactive decay overwhelmingly exceeds that from cocoon-induced shock heating. This can be demonstrated with a simple order-of-magnitude estimate.
The thermal energy deposited by the cocoon is given by $\eta_{\rm coc} E_{\rm coc,th}$, where $E_{\rm coc,th} \simeq 2 L_j t_{\rm bo}$, with the jet breakout time $t_{\rm bo} \simeq 0.4\,\mathrm{s}$. We adopt a cocoon thermalization efficiency of $\eta_{\rm coc} \simeq 0.5$ by following \citet{Hamidani20}.
 For the wide-jet model (HE-WJ) with an opening angle of $\theta_j = 18^\circ$, and adopting a jet power $L_j \simeq L_{\rm iso}\theta_j^2/4 \simeq 1.2 \times 10^{49}\,\mathrm{erg\,s^{-1}}$, we obtain a cocoon thermal energy of $E_{\rm coc,th} \sim 10^{44}\,\mathrm{erg}$ at $t\sim$1 hour. 

By contrast, the energy released through radioactive decay of freshly synthesized $r$-process nuclei is far larger. The specific heating rate approximately follows $\dot{E} \propto t^{-1.3}\,\mathrm{erg\,s^{-1}\,g^{-1}}$ \citep{Metzger10}. For an ejecta mass of $M_{\rm ej} = 0.01\,M_\odot$, the total radioactive energy deposited within the first hour is $E_{\rm rad} \sim 10^{48}\,\mathrm{erg}$, assuming a thermalization efficiency of $\eta \simeq 0.5$. Thus, radioactive heating exceeds the cocoon contribution by roughly three to four orders of magnitude, consistent with previous studies \citep[e.g.,][]{Klion21}.

Given that radioactive heating dominates the energy budget, we neglect any contribution from jet-induced shock heating and include only radioactive heating. The thermalization efficiency of the decay products is computed using the analytic prescription of \ct{Barnes16}. We include the radioactive energy deposition prior to the initial epoch by accounting for its degradation due to adiabatic cooling.
\begin{figure}[t]
\begin{center}
\includegraphics[width=\linewidth]{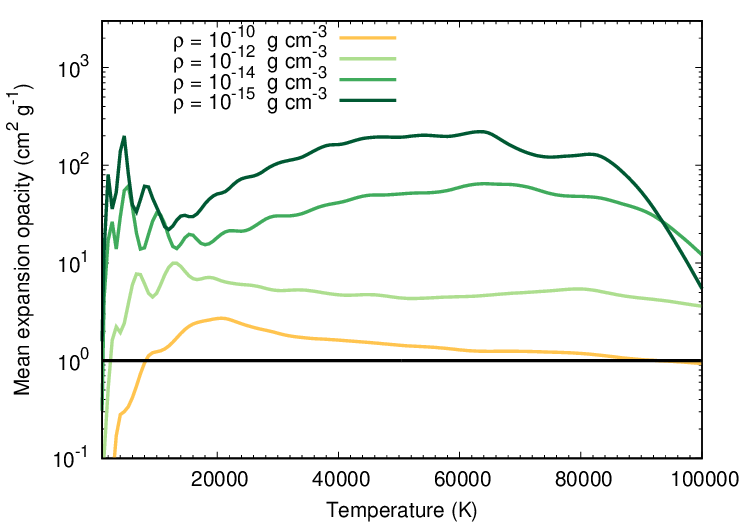}
\end{center}
\caption{Opacity evolution with temperature for different densities at $t = 0.1$ days taken from \ct{Smaranikab20, Smaranikab24}.
The black line shows the gray opacity adopted by most earlier works (e.g., \cta{Klion21}).}
\label{fig:kap}
\end{figure}


\begin{figure*}[t]
  \centering
  \setlength{\tabcolsep}{10pt}

  \begin{tabular}{cc}
    \begin{minipage}[c]{0.55\textwidth}
      \centering
      \includegraphics[width=\linewidth]{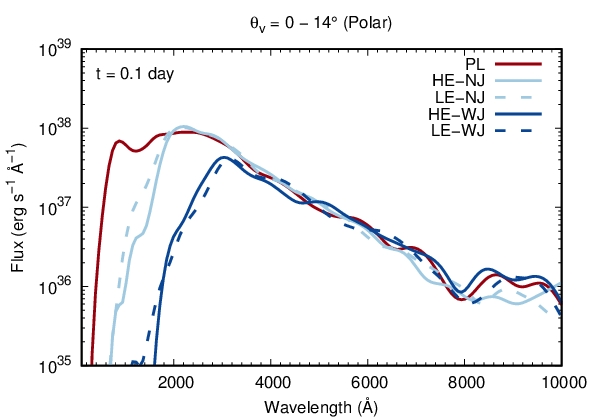}
    \end{minipage}
    &
    \begin{minipage}[t]{0.40\textwidth}
      \centering
      \includegraphics[width=\linewidth]{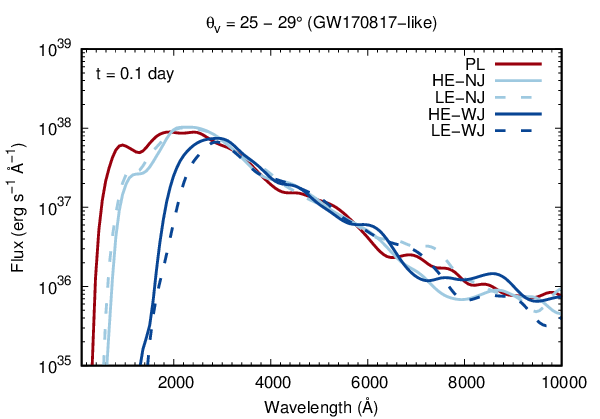}\\[6pt]
      \includegraphics[width=\linewidth]{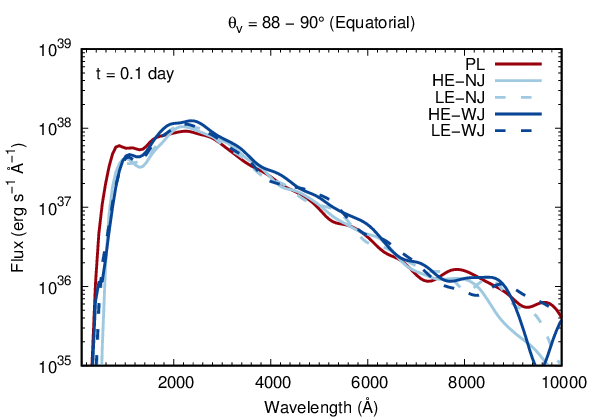}
    \end{minipage}
  \end{tabular}

  \caption{Kilonova spectra for different models at viewing angles: polar ($\theta_{\rm v}=0$–$14^\circ$, top left), GW170817-like ($\theta_{\rm v}=25$–$29^\circ$, top right), and equatorial ($\theta_{\rm v}=88$–$90^\circ$, bottom right). The spectral peak shifts towards longer wavelengths for the jet-interacted ejecta, especially for near-polar viewing angles. For the equatorial viewing angle, the differences almost vanish.}

\label{fig:spec}
\end{figure*}
\subsection{Ejecta physical state} \label{sec:pc_ej}
\begin{figure*}[t]
\centering
\setlength{\tabcolsep}{2pt}%
\renewcommand{\arraystretch}{0}%
\captionsetup{font=small, width=\textwidth, belowskip=0pt, aboveskip=4pt}
\begin{tabular}{@{}cc@{}}
\includegraphics[width=0.45\textwidth]{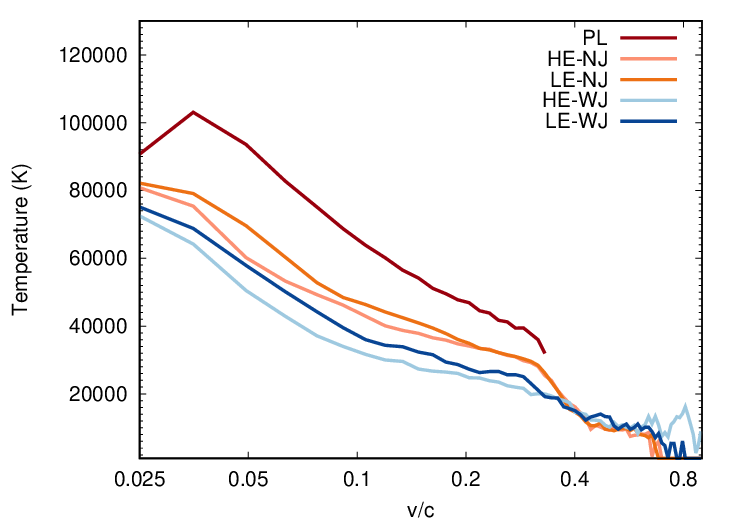} &    
\includegraphics[width=0.45\textwidth]{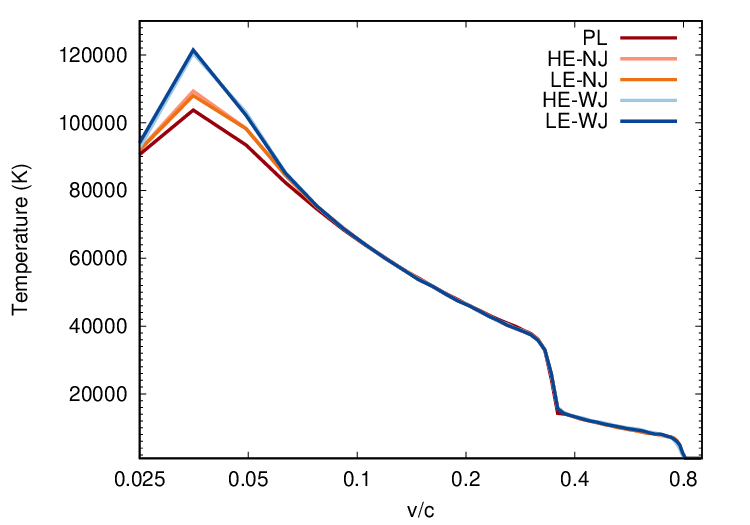}
\end{tabular}
\caption{Temperature profiles for the different models at $t = 0.1$ days are shown along the polar ($\theta_{\rm v}=0$–$14^\circ$, left) and equatorial ($\theta_{\rm v}=88$–$90^\circ$, right) directions. The jet-interacted ejecta exhibit modified temperature structures, reflecting the modified densities caused by the jet interaction, particularly along the polar direction. At high velocities in the polar region, these deviations are the strongest in the wide-jet models, consistent with their more significant modification of the thin outer layer. In contrast, along the equatorial direction, temperature differences are small, except in the innermost ejecta, where minor changes arise from weak density perturbations induced by the jet.}
\label{fig:pc}
\end{figure*}

In the expanding ejecta of a neutron star merger, the physical conditions, such as density and temperature, and consequently the opacity, evolve with time. For instance, the density drops as $\rho(r,t) \propto t^{-3}$ under homologous expansion. Correspondingly, the temperature and opacity evolve.
The Monte Carlo radiative-transfer code used in this work accounts for these effects self-consistently. 
In the following, we briefly outline how the temperature and opacity are determined at each timestep of calculation,
while further details can be found in \citet{Tanaka13, Kawaguchi18, Kawaguchi21}.

The temperature is determined from the photon flux, under the assumption that the electron temperature is equal to the radiation temperature.
The photon intensity is first calculated in individual cells after the photon transfer \ctp{Lucy03, Tanaka13}:
\begin{equation}
    J_{\nu} = \dfrac{1}{4\pi\Delta t V}\sum_{d\nu} \epsilon ds,
\end{equation}
where the summation is over the photon trajectory segment, $ds$, for the photons in the discretized cell of volume $V$, with the comoving energy $\epsilon$, over the timesteps $\Delta t$. This intensity is then integrated over frequency to calculate $<J> = \int J_{\nu} d\nu$, which is related to the temperature as per Stefan-Boltzmann law, $ <J>= \sigma T^4/\pi$. This treatment is adopted from \ct{Kromer09}.

The density and temperature control the ionization state and atomic level populations of the ejecta, and therefore determine the opacity. The opacity influences the photon propagation in the ejecta, hence the light curves and spectra of the kilonova. Here, we mainly discuss the bound-bound opacity, which makes the dominant component in the neutron star merger ejecta in the early time. In the case of supernovae and neutron star mergers, where the matter is expanding with a high velocity and a high velocity gradient, we use the expansion opacity formalism \ctp{Sobolev60, Pinto00} as:
\begin{equation}\label{eqn:kexp}
    \kappa_{\rm{exp}}(\lambda) = \frac{1}{\rho ct}\sum_{l}\frac{\lambda_{l}}{\Delta \lambda}(1 -e^{-\tau_{l}}),
\end{equation}
where $\lambda_{l}$ is the transition wavelength in the wavelength interval of $\Delta \lambda$
and $\tau_{l}$ is the Sobolev optical depth at the transition wavelength, calculated as
\begin{equation}\label{eqn:tau}
    \tau_{l} = \frac{\pi e^{2}}{m_{\rm{e}} c} n_{l}\lambda_{l}f_{l}t.
\end{equation}
Here $f_{l}$ is the strength of transition and $n_{l}$ is the number density of the lower level of the transition,
evaluated by the Boltzmann distribution: 
\begin{equation}
n_l = \frac{g_l}{U(T)} \exp(-E_l/kT),
\end{equation}
where $g_l$ and $E_l$ are the statistical weights and the excited energy of the lower level, respectively, and $U(T)$ is the partition function of the ion.

We perform the radiative-transfer calculations under the assumption of local thermal equilibrium (LTE). Hence, we solve the Saha ionization equation to calculate the ionization fraction. Previous studies \ctp{Kasen13, Smaranikab22} have shown that this approximation is valid for the first few days after the merger. Since our focus is on very early times ($t \leq 1$ day), this treatment is appropriate.

\autoref{fig:kap} shows the Planck mean expansion opacity as a function of temperature for several representative ejecta densities (see \autoref{fig:den2}) at $t\sim0.1$ days. The opacities are calculated for an electron fraction of \(Y_e = 0.30\text{-}0.40\), i.e., the lanthanide-free elemental compositions ($Z = 20 - 56$) over the wavelength range \(\lambda = 100\text{-}35{,}000~\mathrm{\AA}\). 
These data are based on the opacity developed by \citet{Smaranikab20, Smaranikab24}, which incorporate highly ionized (up to XI) \(r\)-process elements, enabling accurate treatment of early-time kilonova emission (starting $t \sim 1$ hour) under high-temperature conditions.

The opacity increases as density decreases, as expected from \ar{eqn:kexp} and \ar{eqn:tau}. For example, at a fixed temperature of $T = 40,000\,\mathrm{K}$, the maximum opacity varies from $\kappa \sim 2\,\mathrm{cm^2\,g^{-1}}$ to $\kappa \sim 200\,\mathrm{cm^2\,g^{-1}}$, depending on the density (\ar{fig:kap}). In the thin outer layer, where the density can be as low as $\rho \lesssim 10^{-14}\,\mathrm{g\,cm^{-3}}$ (\ar{fig:den2}), the opacity can therefore be substantially higher than the values typically assumed in previous studies (e.g., $\kappa \sim 1\,\mathrm{cm^2\,g^{-1}}$; \citealt{Klion21}).

The variation of opacity with temperature is more complicated. With a change in temperature for a particular density, the ionization changes. For different ions, the energy level distribution is different. Depending on the energy level structure in the particular ions, the opacity values are varied \ctp{Smaranikab24}. For more details on the temperature variation of opacity, see \ct{Tanaka20a, Smaranikab24}.

\section{Results} \label{sec:res}
\subsection{Spectra} \label{subsec:spec}

\begin{figure*}[t]
  \begin{tabular}{c}
 
   \begin{minipage}{0.5\hsize}
      \begin{center}        
        \includegraphics[width=\linewidth]{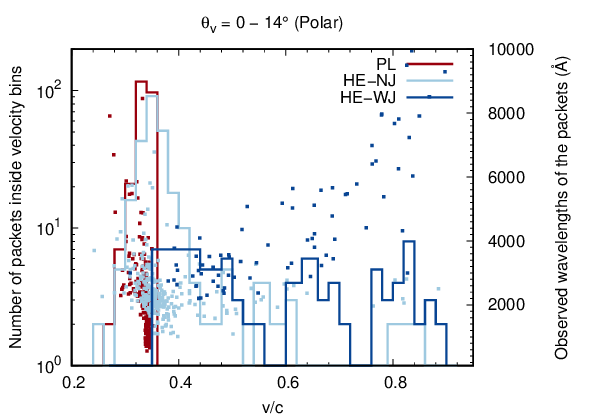}
      \end{center}
    \end{minipage}

    \begin{minipage}{0.5\hsize}
      \begin{center}        
        \includegraphics[width=\linewidth]{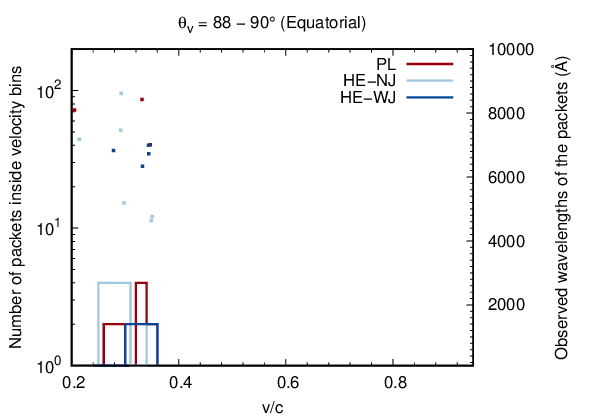}
      \end{center}
    \end{minipage}

\end{tabular}
\caption{Distribution of the ejecta velocities at which photon packets undergo their final interaction before escaping. The points represent the corresponding wavelengths of escaping photon packets (right‑hand axis). The particular colors represent individual models. When a jet interacts with the ejecta (models NJ and WJ), the ejecta produces a thin layer in front of the ejecta. Consequently, the last‑scattering sites of the photons lie at relatively higher velocities than in the PL model with no jet-ejecta interaction. Also, the wavelengths of the emitted photons are on the longer wavelength side for the jet-interacted ejecta models.}
\label{fig:lss}
\end{figure*}
\begin{figure*}[t]
  \centering
  \setlength{\tabcolsep}{10pt}

  \begin{tabular}{cc}
    \begin{minipage}[c]{0.55\textwidth}
      \centering
      \includegraphics[width=\linewidth]{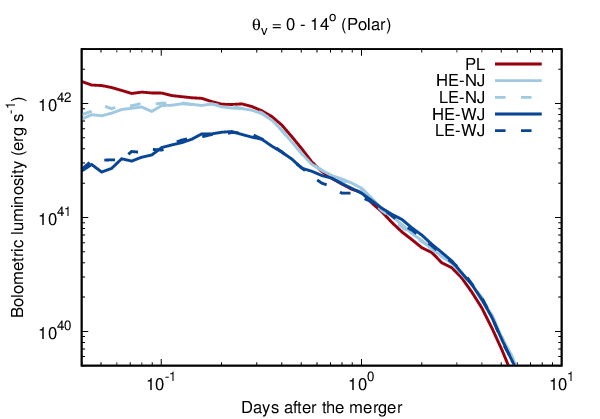}
    \end{minipage}
    &
    \begin{minipage}[t]{0.40\textwidth}
      \centering
      \includegraphics[width=\linewidth]{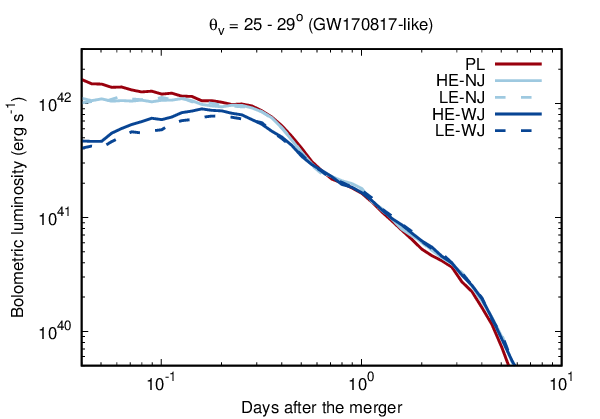}\\[6pt]
      \includegraphics[width=\linewidth]{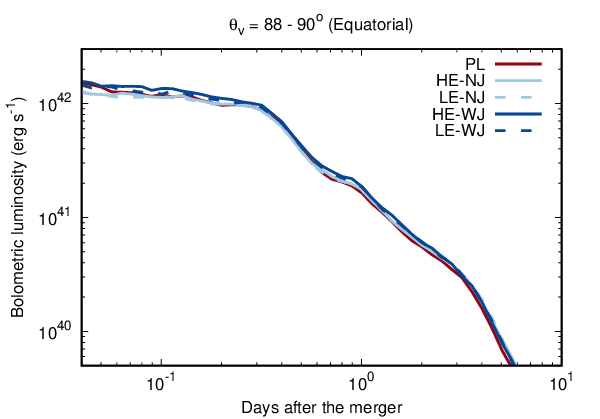}
    \end{minipage}
  \end{tabular}

  \caption{Bolometric light curves for different models shown in \ar{tab:model}, with different panels representing different viewing angles. The presence of the jet makes the light curves dimmer, with the signature most pronounced in the polar direction ($\theta_{\rm v} = 0 - 14^{\circ}$). The wider jet (WJ) models show a stronger effect on light curves.}
  \label{fig:lbol}
\end{figure*}

\ar{fig:spec} shows the spectra at $t = 0.1$ days for the different ejecta models, displayed for three representative viewing-angles measured from the pole: $\theta_{\rm v} = 0$-$14^{\circ}$ (polar), $\theta_{\rm v} = 25$-$29^{\circ}$ (GW170817-like; \citealt{Finstad18, Mooley18}), and $\theta_{\rm v} = 88$-$90^{\circ}$ (equatorial).
At this early epoch, all models, independent of the viewing angle, exhibit spectral peaks in the short-wavelength ultraviolet regime ($\lambda \approx 1,800$-$3,000\,\text{\AA}$). However, both the detailed spectral shapes and the absolute flux levels vary significantly with viewing angle and ejecta model.
Note that we focus on the global spectral shape rather than detailed features, as the current atomic data are not sufficiently accurate for the feature identification.

The origin of the spectral peaks at shorter wavelengths can be understood from the temperature structure at $t = 0.1$ days (\ar{fig:pc}). The temperature, shown for both the polar and equatorial directions, peaks near the center and decreases outward in all models, consistent with the higher densities in the inner ejecta. In the polar direction, the temperature at $v = 0.3c$ ranges between $T \approx 20{,}000$--$40{,}000\,\mathrm{K}$ across the models, while the temperatures in the inner layers are higher. Even at higher velocities (e.g., $v \gtrsim 0.4c$), the ejecta cools but can still reach temperatures as high as $T \sim 15{,}000\,\mathrm{K}$ (\ar{fig:pc}). These elevated ejecta temperatures naturally shift the spectral energy distribution toward the ultraviolet at early times, resulting in peaks at shorter wavelengths.

The effect of jet–ejecta interaction on the spectra is strongest for near-polar viewing angles, as shown in \ar{fig:spec}. At a given observing angle, the location of the spectral peak depends primarily on the jet opening angle: models with wider jet produce progressive shifts towards longer wavelengths relative to the PL model. For a polar observer ($\theta_{\rm v}=0$–$14^\circ$), the PL, NJ, and WJ models peak at approximately $\lambda \simeq 1800\,\mathrm{\AA}$, $2200\,\mathrm{\AA}$, and $3000\,\mathrm{\AA}$, respectively. By contrast, models with the same jet opening angle but different jet powers (e.g., HE–NJ versus LE–NJ) show only minor differences, and the peak wavelength remains largely unchanged.

The model-to-model differences diminish as the viewing angle increases. For polar viewing angles, the peak-wavelength separation between the PL and HE–WJ models is about $\Delta\lambda \simeq 1200\,\mathrm{\AA}$, while at a GW170817-like inclination ($\theta_{\rm v}\approx25^\circ$–$29^\circ$), it decreases to $\Delta\lambda \simeq 1000\,\mathrm{\AA}$ (top right panel of \ar{fig:spec}). For equatorial viewing angles, the spectral differences between the PL and jet-interacted models become negligible (bottom right panel of \ar{fig:spec}).

To diagnose the origin of the differences in peak wavelength across models, we examine the physical conditions of the ejecta at the locations where photons undergo their final interaction (last scattering or absorption) before escaping.
\ar{fig:lss} maps the velocity distribution (the histogram) and escape wavelengths (points) of the escaping photon packets.
For simplicity, we consider only the high-energy jet models (HE), as the low-energy (LE) models exhibit similar behavior. Moreover, we show only photons escaping toward the polar and equatorial directions, displayed in the left and right panels of \ar{fig:lss}, respectively.

When a jet interacts with the ejecta (models NJ and WJ), the ejecta produces a thin layer in front of the ejecta. Consequently, the last-scattering sites of the photons lie at relatively higher velocities than in the PL model with no jet-ejecta interaction. For instance, the photon escape in the polar direction happens at around $v_{\rm ej} \sim 0.35c$,
whereas the same for the models NJ and WJ are much higher ($v_{\rm ej} \sim 0.4-0.8c$, \ar{fig:lss}). Although the velocity from where the majority of the photons escape is not largely different for PL and NJ models ($v_{\rm ej} \sim 0.32c$ and $v_{\rm ej} \sim 0.36c$, respectively), it can be as high as $v_{\rm ej} \sim 0.8c$ for WJ model (\ar{fig:lss}).
This is due to higher opacity at thinner layers formed in the jet-interacted models (see \ar{fig:kap}).

In the polar direction, the faster-moving ejecta, which are characterized by lower densities, have cooler temperatures than the dense inner regions because of the reduced total radioactive energy deposition (\ar{fig:pc}). For example, in the wide-jet (WJ) models, the temperature in the thin outer layer at $v_{\rm ej} \sim 0.6c$ reaches only $T \sim 15,000\,\mathrm{K}$ at $t \sim 0.1$ days. In contrast, for the PL model, the temperature at the outer edge of the ejecta from which photons escape (e.g., at $v_{\rm ej} \sim 0.32c$) reaches $T \sim 40,000\,\mathrm{K}$ at the same epoch (\ar{fig:pc}). Consequently, in the jet-interacted models, especially in the wide-jet case, the ejecta layers from which photons escape are significantly cooler, shifting the spectral peak toward longer wavelengths. This effect is also reflected in \ar{fig:lss}, where the escape wavelengths of individual photon packets (points) are more broadly distributed for the wide-jet model.

For the equatorial direction, the physical properties of the ejecta from where the photons are absorbed and re-emitted before finally escaping the ejecta are not significantly different among models (right panel, \ar{fig:lss}). For instance, for the models, PL and HE-WJ, the maximum velocity of the ejecta where the majority of the photons interacted before leaving the ejecta is $v_{\rm ej} \sim 0.35c$.
This is because the ejecta is not significantly affected around this region by the jet (\ar{fig:den1}).
The temperature around the equatorial region for all the different models is the same ($T \sim 40,000$ K, \ar{fig:pc}). This is also evident from the distribution of the wavelengths in the escaped photons. Hence, the differences among the spectra almost completely disappear in the equatorial direction.

\begin{figure}[t]  
      \begin{center}       
       \includegraphics[width=\linewidth]{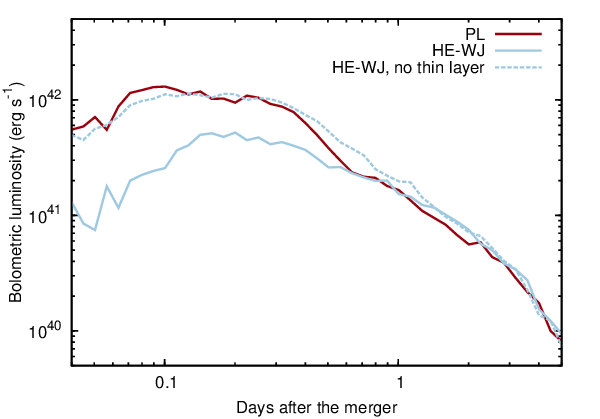}
      \end{center}
  \caption{The bolometric luminosities at the polar viewing angle ($\theta_{\rm v} = 0-14$ degree) for the model HE-WJ with (solid blue) and without (dashed blue) the thin layer created at $v_{\rm ej}>0.35c$ by the jet-ejecta interaction. By artificially removing the thin layer, the curves are almost identical to the light curves calculated using the ejecta with no jet interaction (PL, red curve).}
  \label{fig:lbol_tl}
\end{figure}


\begin{figure*}[t]
\centering
\setlength{\tabcolsep}{2pt}      
\renewcommand{\arraystretch}{0.8}

\begin{tabular}{ccc}
\includegraphics[width=0.32\linewidth]{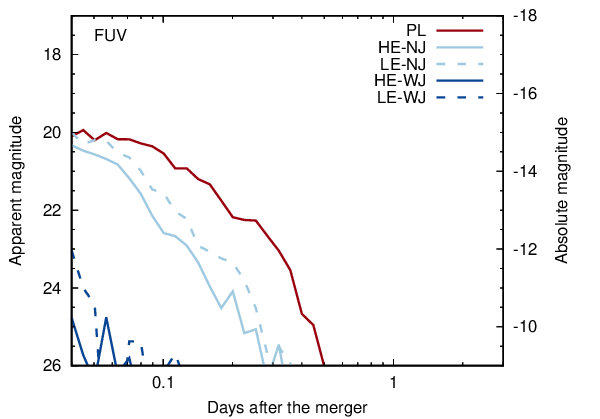} &
\includegraphics[width=0.32\linewidth]{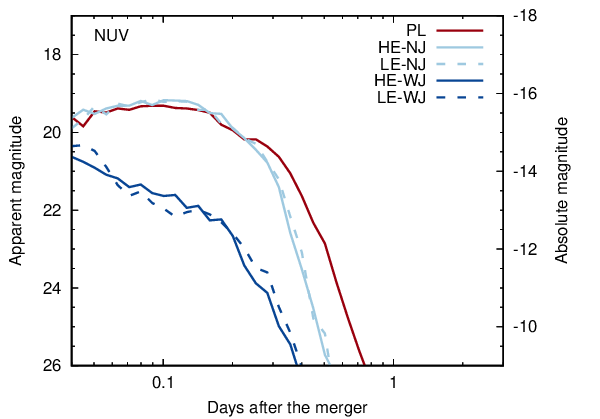} &
\includegraphics[width=0.32\linewidth]{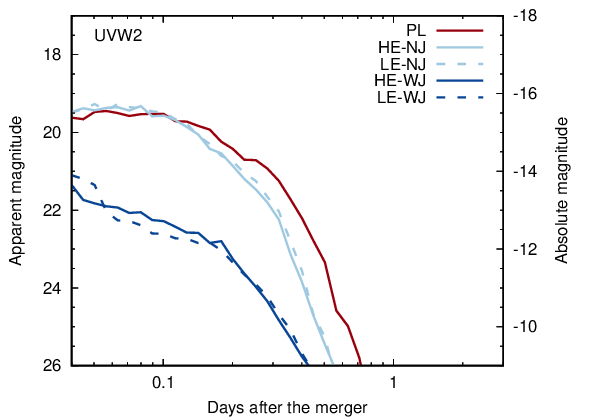} \\[-2pt]

\includegraphics[width=0.32\linewidth]{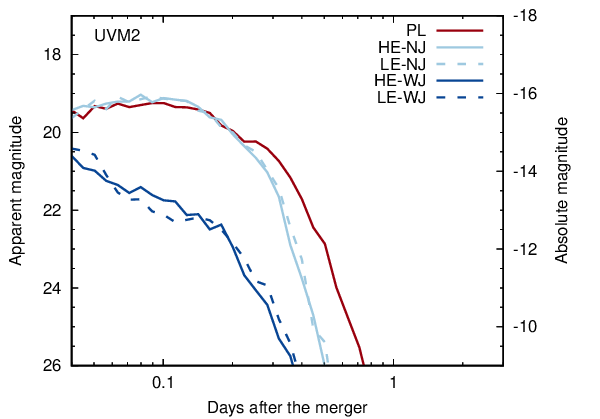} &
\includegraphics[width=0.32\linewidth]{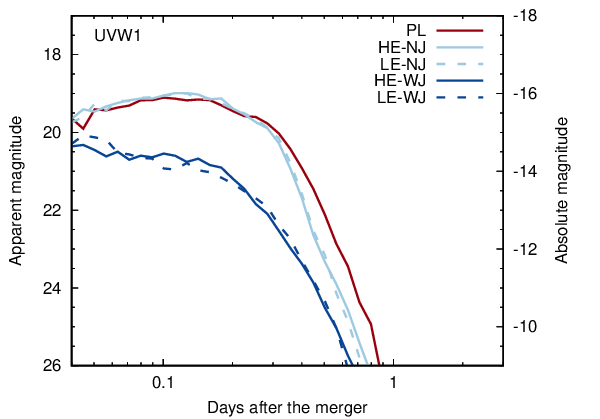} &
\includegraphics[width=0.32\linewidth]{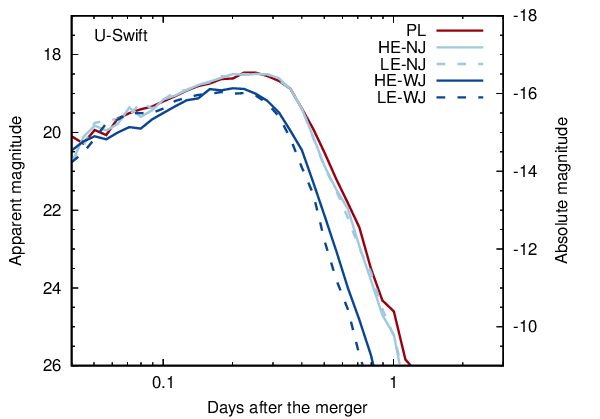} \\[-2pt]

\includegraphics[width=0.32\linewidth]{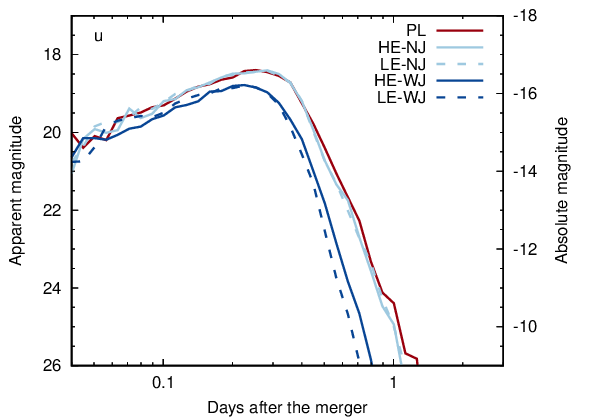} &
\includegraphics[width=0.32\linewidth]{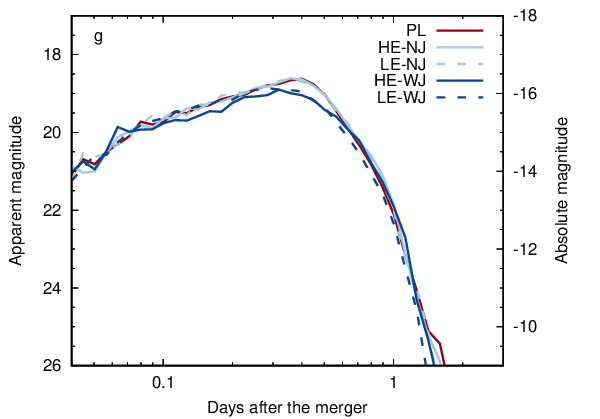} &
\includegraphics[width=0.32\linewidth]{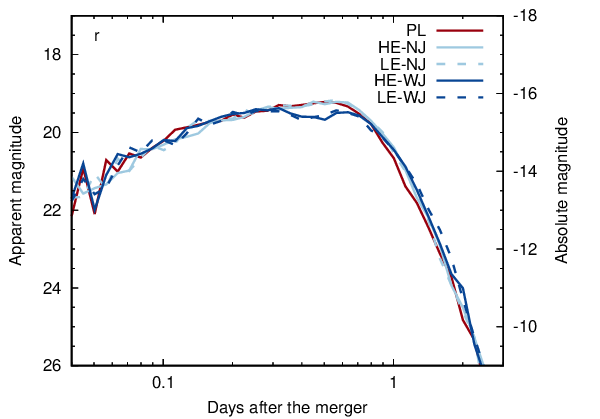} \\
\end{tabular}

\caption{
The multi-color light curves at polar viewing angles ($\theta_{\rm v}=0$-$14^{\circ}$) for different models at 100~Mpc. Each panel shows distinct ultraviolet and optical filters, including \textit{UVEX}: FUV ($\lambda_{\rm Band}= 1,390–1,900\, \rm \AA$) and NUV ($\lambda_{\rm Band} = 2,030–2,700\, \rm \AA$; calculated using v0.6.1 of the UVEX ETC);
\textit{Swift}: UVW2 ($\lambda_{\rm Band}= 1,597 - 3,480\, \rm \AA$), UVM2 ($\lambda_{\rm Band}= 1,699 - 2,964	\, \rm \AA$), UVW1 ($\lambda_{\rm Band}= 1,614 - 4,730 \, \rm \AA$), U ($\lambda_{\rm Band}= 3,018 - 4,781\, \rm \AA$), $u$ ($\lambda_{\rm Band}= 3,206 - 4,081\, \rm \AA$), $g$ ($\lambda_{\rm Band}= 3,876 - 5,665\, \rm \AA$), and $r$ ($\lambda_{\rm Band}= 5,377 - 7,055\, \rm \AA$). The imprint of jet-ejecta interaction is most pronounced in the UV, U-\textit{Swift}, and $u$-bands.
}
\label{fig:mag}
\end{figure*}

\subsection{Light curves} \label{subsec:bol_lc}
To examine the temporal evolution of the kilonova emission, we present the bolometric light curves for all models in \ar{fig:lbol}, with different panels corresponding to different viewing angles. The presence of a jet suppresses the bolometric luminosity at early times ($t \lesssim 1$ day), with the strongest effect occurring for polar viewing angles. For example, at $t \sim 0.1$ days, the wide-jet model (HE–WJ) is fainter by a factor of $\sim 2.5$ in bolometric luminosity compared to the no-jet (PL) model when viewed from the polar direction. This contrast decreases with increasing viewing angle: at a GW170817-like inclination, it is reduced to a factor of $\sim 1.7$, and it becomes negligible for equatorial observers (\ar{fig:lbol}), consistent with the spectra seen in \ar{fig:spec}.

The fainter polar light curves can be attributed to the presence of a thin outer layer oriented toward the polar direction (\ar{fig:den2}). A similar thin-layer effect has been reported for simpler ejecta configurations by \citet{Kasen17} and \citet{Smaranikab24}. In our models, this thin layer is produced by the interaction between the jet and the ejecta and is characterized by low density and a lower temperature than the inner layers, leading to enhanced opacity (\ar{fig:kap}). As a result, at early times most photons escape from this optically thick outer layer (\ar{fig:lss}), yielding fainter emission in the polar direction. No comparable thin layer forms in the equatorial direction, which explains why the equatorial light curves remain largely similar across different models.

To further elaborate on this interpretation, we recompute the light curves for the wide jet-interacted ejecta model (HE–WJ) after artificially removing the thin outer layer beyond the velocity edge of the power-law ejecta ($v_{\rm ej} > 0.35c$). With this layer removed, the polar light curve (dashed blue curve in \ar{fig:lbol_tl}) becomes nearly indistinguishable from that of the no-jet model (red curve in \ar{fig:lbol_tl}), particularly at early times, demonstrating that the thin outer layer is responsible for the suppressed polar emission.

Note that we have assumed the homogeneous abundance distribution to calculate the light curves. But in reality, the equatorial ejecta will have lower $Y_{e}$ than assumed. Hence, this assumption can affect values of the light curves and spectra in the equatorial direction, although it does not alter the relative behavior of models with and without a jet.

The early emission is more sensitive to the jet opening angle than to the jet power. The differences in the light curves between the wide and narrow jet (WJ and NJ) models are much more than the high-energy (HE) and low-energy (LE) jet models, as also seen in the spectra. This is because the differences in the light curves stem from the thin outer layer formed during jet propagation. The jet power itself has only a weak influence on the formation of this layer, because it enters the jet propagation equations linearly (as $L_j$ or $L_{\rm iso}$), whereas the jet opening angle enters quadratically, scaling as $\theta_j^2$. As a result, wide jets (WJ) sweep up and thermalize a much larger fraction of the ejecta, producing a more massive thin layer in comparison to narrow jets (NJ). This explains why HE and LE models yield nearly identical light curves and spectra, whereas WJ and NJ models show pronounced differences.

As time progresses, the light curves of the jet-interacted models (NJ and WJ) converge toward those of the no-jet model (PL), even for polar viewing angles. This occurs because the photosphere recedes deeper into the ejecta, where differences in the density and temperature structures between models become smaller, causing the signatures in the light curves to disappear. We note that our models adopt a relatively low ejecta mass appropriate for the disk-wind component of binary neutron star mergers through which the jet propagates \citep{Hamidani23a, Hamidani23b}. For larger ejecta masses, this effect would persist to later times owing to the longer diffusion timescale.

We also compute multi-color light curves (\ar{fig:mag}) to quantify the impact of jet-ejecta interaction on the broadband emission. We focus primarily on the polar viewing angle, where differences in both spectra and light curves are most pronounced. We present results in several representative ultraviolet and optical filters  (\textit{UVEX} NUV, FUV; \textit{Swift} UVW2, UVW1, UVM2, U-\textit{Swift}, $u$-, $g$-, and $r$-bands), as the early-time emission peaks at shorter wavelengths (\ar{fig:spec}). Magnitudes are shown only for the polar orientation, where the observational signatures of the jet-ejecta interaction are maximally visible, consistent with the trends seen in the spectra and bolometric light curves.

We find that the impact of the jet-ejecta interaction is strongest in the ultraviolet and $u$-bands, while it becomes negligible in the $g$- and $r$-bands. For a given UV or $u$-band, the PL model is consistently the brightest, whereas the WJ models are the faintest. In the UVW2 band, for example, the UV luminosity for PL model reaches a peak of $\simeq 19.5$ mag at $t \simeq 0.15$ days for a source at 100 Mpc, while the WJ models produce fainter emission at the same epoch, $\sim 22$ mag. 
Finally, we find that models with lower jet energies (LE) show only minor differences relative to their high-energy (HE) counterparts.

Among the ultraviolet bands, the jet–ejecta interaction produces the strongest signatures in the \textit{UVEX} NUV, \textit{Swift} UVW2, and UVM2 filters. We therefore recommend prioritizing follow-up observations in these bands to detect the imprint of the jet–ejecta interaction. The differences in the UV and $u$-band magnitudes persist only for a short period, $t \lesssim 1$ day, implying that rapid follow-up after discovery is essential. Looking ahead, early-time ultraviolet observations with forthcoming facilities such as \textit{ULTRASAT} (limiting magnitude $\sim 22.4$ mag for 900\,s exposure; \citealt{Sagiv14}) and \textit{UVEX} (limiting magnitude $\sim 25$ mag for 900\,s exposure; \citealt{Kulkarni21}) will enable systematic measurements of early light curves, making it possible to robustly identify the observational signatures of jet–ejecta interaction.

Thus far, we have focused on the multi-color light curves viewed along the polar direction. To determine how far from the pole the signatures of jet–ejecta interaction remain observable, we examine the dependence of the \textit{Swift} UVW2-band magnitude on viewing angle at $t = 0.1$ days and $t = 0.4$ days (solid and dashed curves in \ar{fig:magt_01}) for the HE–WJ model. We find that the jet signature remains clearly detectable in the UVW2 band out to viewing angles of $\theta_{\rm v}\sim 60^\circ$. This wide angular visibility allows the jet–ejecta interaction to be identified even when the afterglow, which is confined to narrower polar angles, is weak or undetectable. More on the observational implications of this result is discussed in the following section.

\begin{figure}[t]
  
      \begin{center}
       
       \includegraphics[width=\linewidth]{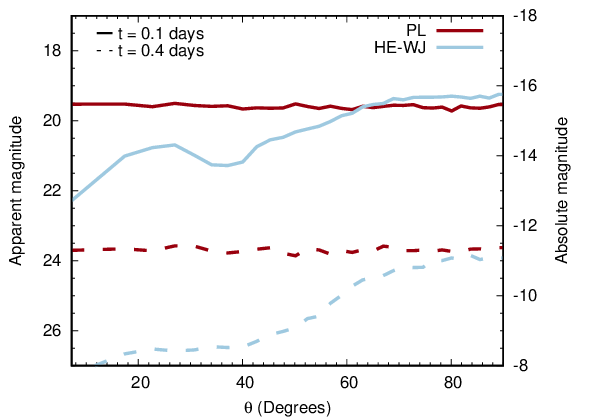}
      \end{center}
  \caption{Variation of the UVW2-band apparent magnitude, shown for a source at a distance of 100\,Mpc, with viewing angle at $t=0.1$ days (solid curves) and $t=0.4$ days (dashed curves). The signature of the jet–ejecta interaction remains clearly detectable out to viewing angles of $\theta_{\rm v} \simeq 60^\circ$.
}
  \label{fig:magt_01}
\end{figure}

\section{Discussion} \label{sec:disc}
In this section, we compare our results with the other works (\ar{subsec:compr}), and we discuss the conditions under which the jet signature in the earliest kilonova is visible beyond the afterglow resulting from the jet and the interstellar medium (ISM) interaction (\ar{subsec:ag}).

\subsection{Comparison with previous works} \label{subsec:compr}
To place our results in the context of previous studies, we first consider the study by \citet{Klion21}, who have used the axisymmetric relativistic hydrodynamic models of \citet{Duffell18} with the Monte Carlo radiative transfer code \textsc{SEDONA} \citep{Kasen06}. Their work, similar to ours, examines how a relativistic jet affects kilonova ejecta and alters the emergent light curves. The principal difference between our works lies in the opacity treatment. While \citet{Klion21} adopted a single, gray value of opacity ($\kappa = 1\ \rm{cm^{2}\,g^{-1}}$) to perform the radiative transfer simulations, we apply wavelength-, temperature-, and time-dependent opacities derived from detailed atomic data \ctp{Smaranikab20, Smaranikab22, Smaranikab24}.

The effect of using constant gray opacities is clearly demonstrated by the different predictions obtained from the two works. For example, \citet{Klion21} found that polar light curves are brighter than equatorial ones, attributing this to the fact that the ejecta, being modified by the jet, expose the hotter inner ejecta along polar viewing angles. This is contradictory to our results.

The discrepancy can be traced back to the treatment of the opacities in the two works. As \ct{Klion21} have used a gray opacity with relatively lower constant value ($\kappa = 1\,\rm cm^2\,g^{-1}$), they find their photosphere residing near the hotter core in the polar direction. 
In contrast, in our models, the photosphere resides in a thin outer layer of the ejecta, as the opacity in this region is significantly enhanced by the low densities in the outermost layers (\ar{fig:kap}). As a result, photons predominantly escape from this outer thin layer in jet-interacted ejecta (\ar{fig:lss}), rather than from the hotter inner core along the polar direction at early times.
As the thin layer has relatively low density, the overall energy deposition is lower than that of the highly dense bulk ejecta, causing the light curves in the polar direction to be fainter than those in the equatorial direction. The contrast underscores the sensitivity of kilonova observables to both the hydrodynamic structure and the detailed opacity.

To assess the impact of using wavelength-dependent opacities, we recalculated one of our models assuming a gray opacity of $\kappa = 0.8\,\mathrm{cm^{2}\,g^{-1}}$ (see \ar{fig:lbol_grayop}). This value was recommended by \ct{Smaranikab25} as providing the best match to the early-time bolometric light curves of light $r$-process kilonovae \citep{Smaranikab24}. However, as also emphasized by \ct{Smaranikab24}, such a gray-opacity approximation fails to reproduce the multi-color light curves.

As shown in \ar{fig:lbol_grayop}, both the peak luminosity and the rise time of the polar (blue) and equatorial (red) light curves change substantially when switching from wavelength-dependent opacities (solid lines) to a gray opacity (dashed lines). In the gray-opacity case, the magnitude difference between the polar and equatorial light curves is significantly reduced compared to the wavelength-dependent treatment. Moreover, the gray-opacity model produces a trend in which the polar light curve becomes brighter than the equatorial one at $t \gtrsim 0.25$ days, consistent with the results of \ct{Klion21}. This behavior is not present when detailed opacities are used, demonstrating that reliance on gray opacities can be misleading when interpreting the signatures of jet–ejecta interaction.

Next, we compare our results with those of \cite{Nativi21} and \ct{Shrestha23}. \citet{Nativi21} model the hydrodynamic interaction between a narrow relativistic jet (opening angle $\theta_{j} = 5^{\circ}$ and kinetic energies of $E_{k} = 10^{49}\,\rm{erg}$ and $E_{k} = 10^{51}\,\rm{erg}$) and inject the jet in the neutrino‐driven wind produced by a long‑lived merger remnant; the wind profile is taken from \citet{Perego14}. They post‑process the outflow with the Monte‑Carlo radiative‑transfer code \texttt{POSSIS} \citep{Bulla19}, adopting analytic opacity prescriptions developed based on \citet{Tanaka18}. Their synthetic light curves reproduce the trend reported by \citet{Klion21}, i.e., emission viewed along the polar axis is brighter than that seen from the equatorial plane, a consequence of the jet-ejecta interaction in the polar direction, altering the density structure in the polar direction, exposing the high-temperature inner ejecta.

\citet{Shrestha23} refine the framework developed by \citet{Nativi21} by recalculating the kilonova emission with the wavelength‑dependent opacities of \citet{Tanaka20a}. While this upgrade improves the results by showing the trend of equatorial light curves being brighter than the polar ones, the adopted opacity up to ion IV is not suitable to be used for the high-temperature ejecta at the early time ($\lesssim$\,1 day after merger). Hence, our calculations give a more accurate estimate for the earliest signal, especially in the ultraviolet wavelengths, relevant for the future satellites, such as \textit{ULTRASAT} and \textit{UVEX}.

\begin{figure}[t]  
      \begin{center}              
       \includegraphics[width=\linewidth]{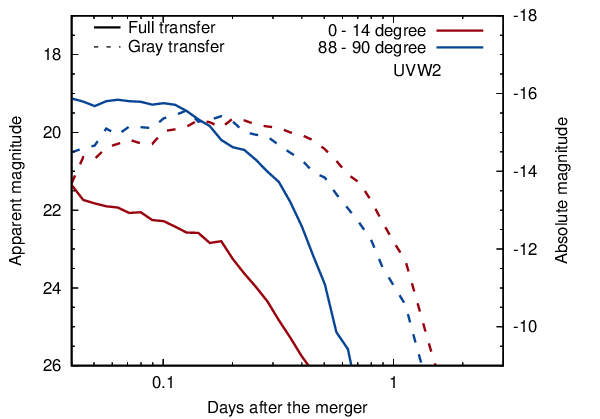}
      \end{center}
  \caption{The UVW2 light curves for the HE–WJ model at polar ($\theta_{\rm v}=0$–$14^\circ$) and equatorial ($\theta_{\rm v}=88$–$90^\circ$) viewing angles for a source at 100 Mpc. We compare light curves calculated using wavelength-dependent opacities (solid curves) and gray opacity (dashed curves) with a value of $\kappa = 0.8\,\mathrm{cm^2\,g^{-1}}$. In the gray-opacity case, the polar light curve becomes brighter than the equatorial one at $t \gtrsim 0.2$ days. This is not physical, and is a consequences of using gray opacity, as can be understood from the results with detailed opacity.}
\label{fig:lbol_grayop}
\end{figure}

\begin{figure*}[t]
  \begin{tabular}{c}
 
   \begin{minipage}{0.5\hsize}
      \begin{center}
       
        \includegraphics[width=\linewidth]{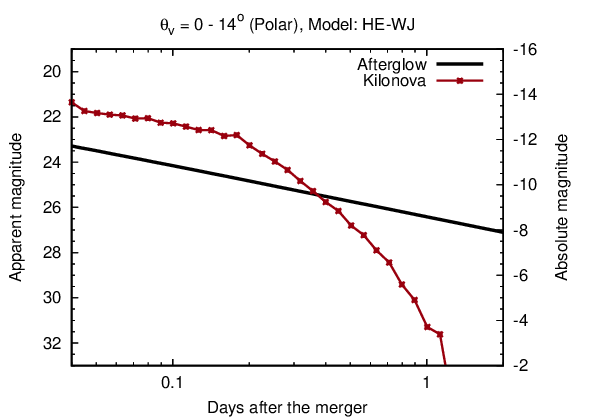}
      \end{center}
    \end{minipage}

 \begin{minipage}{0.5\hsize}
      \begin{center}
        \includegraphics[width=\linewidth]{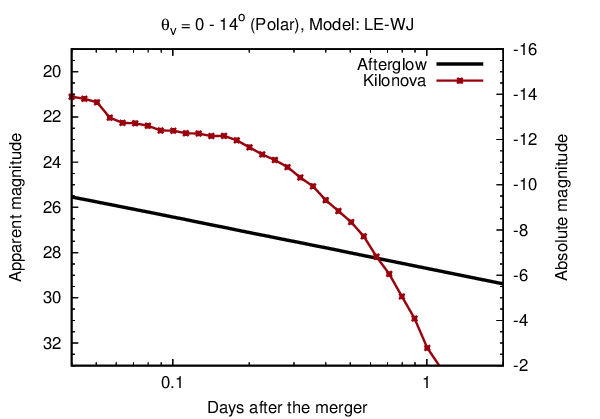}        
      \end{center}
    \end{minipage}
          
\end{tabular}
  \caption{Afterglow emission in the UVW2 band for a structured Gaussian jet observed at a near-polar viewing angle ($\theta_{\rm v}=14^\circ$). We adopt isotropic equivalent jet luminosities of $L_{\rm iso}=10^{50}\,\mathrm{erg\,s^{-1}}$ (left) and $5\times10^{50}\,\mathrm{erg\,s^{-1}}$ (right), with the jet opening angle fixed at $\theta_j=18^\circ$, corresponding to the LE–WJ and HE–WJ jet models listed in \ar{tab:model}. Other afterglow parameters are adopted from \ct{Troja18b}: $\theta_{\rm wing}=31.5^\circ$, $p=2.155$, $\log_{10}(\epsilon_e)=-1.22$, $\log_{10}(\epsilon_B)=-3.38$, $\xi_{\rm N}=1.0$, and $d_L=100$\,Mpc and ISM densities of $\log_{10}(n\,{\rm [cm^{-3}]})=-3.1$. The corresponding kilonova light curves for HE-WJ (left) and LE-WJ (right) models are also shown for comparison in the same bands in the polar direction ($\theta_{\rm v} = 0 - 14^{\circ}$).
}
  
  \label{fig:after_01}
\end{figure*}

\begin{figure*}[t]
 \begin{tabular}{c}
 
   \begin{minipage}{0.5\hsize}
      \begin{center}
        \includegraphics[width=\linewidth]{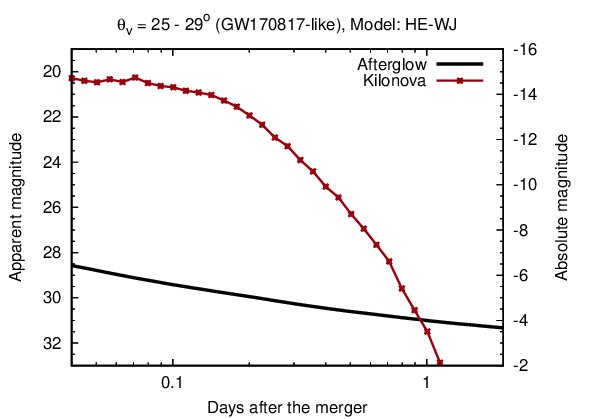}
       
      \end{center}
    \end{minipage}

 \begin{minipage}{0.5\hsize}
      \begin{center}
        \includegraphics[width=\linewidth]{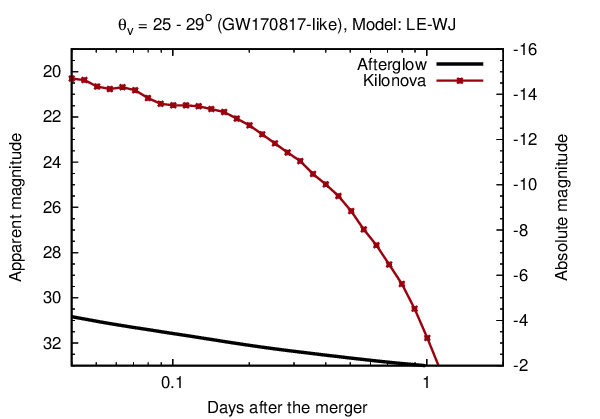}
        
      \end{center}
    \end{minipage}
          
\end{tabular}

\caption{Same as \ar{fig:after_01}, but for an observer at a viewing angle of $\theta_{\rm v}=29^\circ$.} 
\label{fig:after_02}
\end{figure*}
\subsection{Afterglow contamination} \label{subsec:ag}
After the jet breaks out of the ejecta, it eventually propagates into the surrounding ISM. The interaction between the jet and the ISM produces an afterglow spanning a wide range of wavelengths. For an on-axis viewing angle, the afterglow peaks at early times, on a timescale comparable to that of the kilonova emission, and can outshine it. In such cases, the early kilonova emission with the imprint of the jet would not be observable. In contrast, for off-axis viewing angles, the afterglow emission is significantly dimmer, allowing the effects of the jet on the kilonova emission to be detected.

To evaluate the relative contributions of the afterglow and kilonova emission at off-axis viewing angles, we compute the afterglow signal using \texttt{afterglowpy} \citep{Ryan20}. The code calculates afterglow light curves and spectra for different viewing angles based on a set of input parameters, including the jet energy and opening angle, the fractions of post-shock energy in accelerated electrons ($\epsilon_e$) and magnetic fields ($\epsilon_B$), and the interstellar medium (ISM) density ($n$). The microphysical parameters (e.g., $\epsilon_e$) are chosen following the constraints derived from GRB170817A in \citet{Troja18b}.

To compute the afterglow emission, we assume a Gaussian jet structure with a fixed opening angle of $\theta_j = 18^\circ$, chosen to match the wide-jet kilonova model (WJ), for which the jet signature is strongest. We consider both low- and high-energy jets (LE–WJ and HE–WJ models). This is because, although their kilonova emission is nearly identical, the afterglow contribution depends sensitively on the jet energy. We therefore adopt two jet luminosities, $L_{\rm iso} = 1 \times 10^{50}\,\mathrm{erg\,s^{-1}}$ for LE–WJ and $L_{\rm iso} = 5 \times 10^{50}\,\mathrm{erg\,s^{-1}}$ for HE–WJ.

In \ar{fig:after_01}, we show the comparison between the afterglow at $\theta_{v} = 14^{\circ}$ and the kilonova emission for the HE–WJ and LE–WJ models in the polar direction ($\theta_{v} = 0 - 14^{\circ}$) in the \textit{Swift} UVW2 band. In both models, the kilonova is initially brighter than the afterglow during the first few hours. In the HE–WJ case, where the afterglow is stronger, the kilonova remains visible above the afterglow for approximately the first $\sim 10$ hours, after which the afterglow dominates. In the LE–WJ model, the weaker afterglow allows the kilonova to remain visible for a longer period of about $\sim 17$ hours. In the on-axis viewing angle, the afterglow will be brighter by $\sim$ 2 mag, as we found from a crude estimate by using \textit{afterglowpy}. However, even in such a case, the early kilonova will be detectable in the earliest epochs ($t\sim 7$ hours, \ar{fig:after_01}, \ar{fig:after_02}).

At larger viewing angles, the afterglow becomes much fainter because its emission declines more rapidly with increasing angle than the kilonova emission. Consequently, for a GW170817-like viewing angle and the ISM density, the signature of the jet–ejecta interaction in kilonova is clearly detectable in the UV band during the first day, particularly for jets with wide opening angles (\ar{fig:after_02}).

\section{Conclusion} \label{sec:conc}
We present the first radiative-transfer simulations of jet-interacted binary neutron star merger ejecta that incorporate detailed, wavelength-dependent opacities developed in \ct{Smaranikab20, Smaranikab24}, which are suitable for early times ($t \sim 1$ hour after merger). 
At a fixed temperature of $T = 40{,}000\,\mathrm{K}$, representative of the outer edge of the bulk ejecta at $t = 0.1$ days, the maximum opacity for a lanthanide-free composition ($Z = 20$--56) spans a wide range, from $\kappa \sim 2\,\mathrm{cm^2\,g^{-1}}$ to $\kappa \sim 200\,\mathrm{cm^2\,g^{-1}}$, depending on the density (\ar{fig:kap}).
Using this detailed opacities, we investigate how jet–ejecta interaction affects kilonova light curves across a range of jet properties, varying the jet opening angle between $\theta_j = 6.8^\circ$ and $18^\circ$ (NJ and WJ models, \ar{tab:model}) and the isotropic equivalent jet luminosity of $L_{\rm iso} = 10^{50}$ and $5\times10^{50}\,\mathrm{erg\,s^{-1}}$ (LE and HE models, \ar{tab:model}), using hydrodynamic simulations from \ct{Hamidani23a, Hamidani23b}.

Our study shows that the kilonova spectra at $t = 0.1$ days (\ar{fig:spec}) shift the spectral peak toward longer wavelengths in the jet-interacted models (NJ and WJ) compared to the model with no jet-interaction (PL). This effect is most prominent for polar viewing angles ($\theta_{\rm v}=0$–$14^\circ$). This behavior arises because the jet–ejecta interaction forms a thin outer layer in front of the bulk ejecta (\ar{fig:den2}), from which most photons escape at early times (\ar{fig:lss}). This thin layer has a much lower density than the main ejecta, leading to reduced radioactive energy deposition and lower temperatures (\ar{fig:pc}). As a result, the emission originates from cooler material, causing the spectral peak to shift to longer wavelengths. These differences largely disappear for equatorial viewing angles, where the jet has little impact on the ejecta structure.

Our bolometric light curves (\ar{fig:lbol}) show that jet–ejecta interaction suppresses the early-time ($t \lesssim 1$ day) kilonova luminosity, with the strongest effect for polar viewing angles. At $t \sim 0.1$ days, the wide-jet model (HE–WJ) is fainter by a factor of $\sim 2.5$ than the no-jet (PL) model for polar observers, while this difference decreases to $\sim 1.7$ at a GW170817-like inclination and becomes negligible for equatorial views, consistent with the spectra in \ar{fig:spec}. This dimming is caused by the low-density, relatively cooler thin outer layer mainly in the polar direction (\ar{fig:den2}). The physical conditions of this layer (density, temperature) make the opacity at this layer relatively high (\ar{fig:kap}). This causes the photons at the early time to escape mainly from the thin outer layer. In contrast, the ejecta is only weakly affected by the jet in the equatorial direction, and no thin layer is created, leading to nearly indistinguishable light curves for the jet-interacted ejecta in that direction (\ar{fig:lbol}). At later times, all models converge as the photosphere recedes into deeper layers of the ejecta.

We compute multi-color light curves to quantify the impact of jet–ejecta interaction on the broadband emission, focusing on polar viewing angles where the signatures are strongest (\ar{fig:mag}). The results are shown in representative ultraviolet and optical filters, reflecting the fact that the early-time emission peaks at short wavelengths (\ar{fig:spec}). We show the multi-color light curves in the NUV and FUV bands of the upcoming satellite \textit{UVEX}. In addition,  we also show the results in \textit{Swift} UVW2, UVW1, UVM2, and $U$; and the $u$-, $g$-, and $r$-bands. The jet–ejecta interaction produces its largest effects in the ultraviolet and $u$-bands, while differences become negligible in the $g$- and $r$-bands. In all UV and $u$ filters, the PL model is consistently the brightest, and the WJ models are the faintest. For example, in the UVW2 band at 100\,Mpc, the PL model reaches $\simeq19.5$ mag at $t\simeq0.15$\,days, whereas the WJ models are fainter at the same epoch with $\sim22$ mag. The kilonova remains visible even beyond the afterglow at the earliest times for GW170817-like cases, particularly for lower-energy jets and for more off-axis viewing angles. Differences between high- and low-energy jet models (HE versus LE) are small, consistent with their nearly identical kilonova emission.

Among the ultraviolet bands, the strongest and most robust signatures of jet–ejecta interaction appear in the \textit{UVEX} NUV and the \textit{Swift} UVW2 and UVM2 filters, making these the highest-priority bands for follow-up. The differences in the UV and $u$-band light curves persist only for a short period ($t\lesssim1$ day), so rapid observations after discovery are essential. In the near future, forthcoming ultraviolet facilities such as \textit{ULTRASAT} (limiting magnitude $\sim22.4$ mag in 900\,s; \citealt{Sagiv14}) and \textit{UVEX} ($\sim25$ mag in 900\,s; \citealt{Kulkarni21}) will enable early-time UV monitoring for kilonovae, making it possible to robustly identify jet–ejecta interaction in future observations.

\acknowledgments
Numerical simulations presented in this paper were carried out with the computer facility in the Yukawa Institute for Theoretical Physics (YITP),
Kyoto University, Japan. SB thanks Hannah Earnshaw and the UVEX team for providing the filters for the NUV and FUV bands. 
MT is supported by the Grant-in-Aid for Scientific research from JSPS (grant Nos. 23H00127, 23H04894, 23H04891, 21H04997, 23H05432) and the JST FOREST Program (grant No. JPMJFR212Y).
\vspace{5mm}

\bibliography{./references.bib}{}

@ARTICLE{Tanaka20a,
       author = {{Tanaka}, Masaomi and {Kato}, Daiji and {Gaigalas}, Gediminas and
         {Kawaguchi}, Kyohei},
        title = "{Systematic opacity calculations for kilonovae}",
      journal = {\mnras},
         year = 2020,
        month = jun,
       volume = {496},
       number = {2},
        pages = {1369-1392},
        doi = {10.1093/mnras/staa1576},
archivePrefix = {arXiv},
       eprint = {1906.08914},
 primaryClass = {astro-ph.HE},
       adsurl = {https://ui.adsabs.harvard.edu/abs/2020MNRAS.496.1369T}
}

@article{Tanaka13,
	doi = {},
	url = {},
	year = 2013,
	month = {sep},
	publisher = {{IOP} Publishing},
	volume = {775},
	number = {2},
	pages = {113},
	author = {Masaomi Tanaka and Kenta Hotokezaka},
	journal = {ApJ}
}

@ARTICLE{Tanaka18,
       author = {{Tanaka}, Masaomi and {Kato}, Daiji and {Gaigalas}, Gediminas and
        et al.},
      journal = {ApJ},
         year = "2018",
        month = "Jan",
       volume = {852},
       number = {2},
          eid = {109},
        pages = {109},
          doi = {},
archivePrefix = {arXiv},
       eprint = {1708.09101},
 primaryClass = {astro-ph.HE},
       adsurl = {https://ui.adsabs.harvard.edu/abs/2018ApJ...852..109T}
}

@ARTICLE{Kawaguchi18,
       author = {{Kawaguchi}, Kyohei and {Shibata}, Masaru and {Tanaka}, Masaomi},
      journal = {ApJ letter},
         year = "2018",
        month = "Oct",
       volume = {865},
       number = {2},
          eid = {L21},
        pages = {L21},
          doi = {},
archivePrefix = {arXiv},
       eprint = {1806.04088},
 primaryClass = {astro-ph.HE},
       adsurl = {https://ui.adsabs.harvard.edu/abs/2018ApJ...865L..21K}
}

@ARTICLE{Abbott17a,
       author = {{Abbott}, B.~P. and {Abbott}, R. and {Abbott}, T.~D. and {Acernese}, F. and
         {Ackley}, K. and {Adams}, C. and {Adams}, T. and {Addesso}, P. and
         {Adhikari}, R.~X. and {Adya}, V.~B. and {Affeldt}, C. and
         {Afrough}, M. and {Agarwal}, B. and {Agathos}, M. and {Agatsuma}, K. and
         {Aggarwal}, N. and {Aguiar}, O.~D. and {Aiello}, L. and {Ain}, A. and
         {Ajith}, P. and {LIGO Scientific Collaboration} and {Virgo Collaboration}},
        title = "{GW170817: Observation of Gravitational Waves from a Binary Neutron Star Inspiral}",
      journal = {\prl},
         year = 2017,
        month = oct,
       volume = {119},
       number = {16},
          eid = {161101},
        pages = {161101},
          doi = {10.1103/PhysRevLett.119.161101},
archivePrefix = {arXiv},
       eprint = {1710.05832},
 primaryClass = {gr-qc},
       adsurl = {https://ui.adsabs.harvard.edu/abs/2017PhRvL.119p1101A}
}

@ARTICLE{Wanajo14,
       author = {{Wanajo}, Shinya and {Sekiguchi}, Yuichiro and {Nishimura}, Nobuya and
         et al.},
        title = "{Production of All the r-process Nuclides in the Dynamical Ejecta of Neutron Star Mergers}",
      journal = {\apjl},
         year = 2014,
        month = jul,
       volume = {789},
       number = {2},
          eid = {L39},
        pages = {L39},
          doi = {10.1088/2041-8205/789/2/L39},
archivePrefix = {arXiv},
       eprint = {1402.7317},
 primaryClass = {astro-ph.SR},
       adsurl = {https://ui.adsabs.harvard.edu/abs/2014ApJ...789L..39W}
}

@ARTICLE{Barnes16,
       author = {{Barnes}, Jennifer and {Kasen}, Daniel and {Wu}, Meng-Ru and
         {Mart{\'\i}nez-Pinedo}, Gabriel},
        title = "{Radioactivity and Thermalization in the Ejecta of Compact Object Mergers and Their Impact on Kilonova Light Curves}",
      journal = {\apj},
         year = 2016,
        month = oct,
       volume = {829},
       number = {2},
          eid = {110},
        pages = {110},
          doi = {10.3847/0004-637X/829/2/110},
archivePrefix = {arXiv},
       eprint = {1605.07218},
 primaryClass = {astro-ph.HE},
       adsurl = {https://ui.adsabs.harvard.edu/abs/2016ApJ...829..110B}
}

@INPROCEEDINGS{Fernandez14,
       author = {{Fernandez}, Rodrigo and {Metzger}, Brian},
        title = "{Electromagnetic Emission and r-process Nucleosynthesis from Late-Time Winds of Neutron Star Merger Remnant Accretion Disks}",
    booktitle = {AAS/High Energy Astrophysics Division \#14},
         year = 2014,
       series = {AAS/High Energy Astrophysics Division},
        month = aug,
          eid = {304.07},
        pages = {304.07},
       adsurl = {https://ui.adsabs.harvard.edu/abs/2014HEAD...1430407F}
}

@ARTICLE{Hotokezaka13,
       author = {{Hotokezaka}, Kenta and {Kiuchi}, Kenta and {Kyutoku}, Koutarou and
         et al.},
        title = "{Remnant massive neutron stars of binary neutron star mergers: Evolution process and gravitational waveform}",
      journal = {\prd},
         year = 2013,
        month = aug,
       volume = {88},
       number = {4},
          eid = {044026},
        pages = {044026},
          doi = {10.1103/PhysRevD.88.044026},
archivePrefix = {arXiv},
       eprint = {1307.5888},
 primaryClass = {astro-ph.HE},
       adsurl = {https://ui.adsabs.harvard.edu/abs/2013PhRvD..88d4026H}
}

@ARTICLE{Bauswein13,
       author = {{Bauswein}, A. and {Goriely}, S. and {Janka}, H. -T.},
        title = "{Systematics of Dynamical Mass Ejection, Nucleosynthesis, and Radioactively Powered Electromagnetic Signals from Neutron-star Mergers}",
      journal = {\apj},
         year = 2013,
        month = aug,
       volume = {773},
       number = {1},
          eid = {78},
        pages = {78},
          doi = {10.1088/0004-637X/773/1/78},
archivePrefix = {arXiv},
       eprint = {1302.6530},
 primaryClass = {astro-ph.SR},
       adsurl = {https://ui.adsabs.harvard.edu/abs/2013ApJ...773...78B}
}

@ARTICLE{Sekiguchi15,
       author = {{Sekiguchi}, Yuichiro and {Kiuchi}, Kenta and {Kyutoku}, Koutarou and
         {Shibata}, Masaru},
        title = "{Dynamical mass ejection from binary neutron star mergers: Radiation-hydrodynamics study in general relativity}",
      journal = {\prd},
         year = 2015,
        month = mar,
       volume = {91},
       number = {6},
          eid = {064059},
        pages = {064059},
          doi = {10.1103/PhysRevD.91.064059},
archivePrefix = {arXiv},
       eprint = {1502.06660},
 primaryClass = {astro-ph.HE},
       adsurl = {https://ui.adsabs.harvard.edu/abs/2015PhRvD..91f4059S}
}

@ARTICLE{Li98,
       author = {{Li}, Li-Xin and {Paczy{\'n}ski}, Bohdan},
        title = "{Transient Events from Neutron Star Mergers}",
      journal = {\apjl},
         year = 1998,
        month = nov,
       volume = {507},
       number = {1},
        pages = {L59-L62},
          doi = {10.1086/311680},
archivePrefix = {arXiv},
       eprint = {astro-ph/9807272},
 primaryClass = {astro-ph},
       adsurl = {https://ui.adsabs.harvard.edu/abs/1998ApJ...507L..59L}
}

@ARTICLE{Kasen13,
   author = {{Kasen}, D. and {Badnell}, N.~R. and {Barnes}, J.},
    title = "{Opacities and Spectra of the r-process Ejecta from Neutron Star Mergers}",
  journal = {\apj},
archivePrefix = "arXiv",
   eprint = {1303.5788},
 primaryClass = "astro-ph.HE",
     year = 2013,
    month = sep,
   volume = 774,
      eid = {25},
    pages = {25},
      doi = {10.1088/0004-637X/774/1/25},
   adsurl = {http://adsabs.harvard.edu/abs/2013ApJ...774...25K}
}

@ARTICLE{Kasen17,
       author = {{Kasen}, Daniel and {Metzger}, Brian and {Barnes}, Jennifer and
         {Quataert}, Eliot and {Ramirez-Ruiz}, Enrico},
        title = "{Origin of the heavy elements in binary neutron-star mergers from a gravitational-wave event}",
      journal = {\nat},
         year = 2017,
        month = nov,
       volume = {551},
       number = {7678},
        pages = {80-84},
          doi = {10.1038/nature24453},
archivePrefix = {arXiv},
       eprint = {1710.05463},
 primaryClass = {astro-ph.HE},
       adsurl = {https://ui.adsabs.harvard.edu/abs/2017Natur.551...80K}
}

@ARTICLE{Metzger10,
       author = {{Metzger}, B.~D. and {Mart{\'\i}nez-Pinedo}, G. and {Darbha}, S. and
         {Quataert}, E. and {Arcones}, A. and {Kasen}, D. and {Thomas}, R. and
         {Nugent}, P. and {Panov}, I.~V. and {Zinner}, N.~T.},
        title = "{Electromagnetic counterparts of compact object mergers powered by the radioactive decay of r-process nuclei}",
      journal = {\mnras},
         year = 2010,
        month = aug,
       volume = {406},
       number = {4},
        pages = {2650-2662},
          doi = {10.1111/j.1365-2966.2010.16864.x},
archivePrefix = {arXiv},
       eprint = {1001.5029},
 primaryClass = {astro-ph.HE},
       adsurl = {https://ui.adsabs.harvard.edu/abs/2010MNRAS.406.2650M}
}

@ARTICLE{Utsumi17,
       author = {{Utsumi}, Yousuke and {Tanaka}, Masaomi and {Tominaga}, Nozomu and {Yoshida}, Michitoshi and et al.},
        title = "{J-GEM observations of an electromagnetic counterpart to the neutron star merger GW170817}",
      journal = {\pasj},
         year = 2017,
        month = dec,
       volume = {69},
       number = {6},
          eid = {101},
        pages = {101},
          doi = {10.1093/pasj/psx118},
archivePrefix = {arXiv},
       eprint = {1710.05848},
 primaryClass = {astro-ph.HE},
       adsurl = {https://ui.adsabs.harvard.edu/abs/2017PASJ...69..101U}
}

@ARTICLE{Pinto00,
       author = {{Pinto}, Philip A. and {Eastman}, Ronald G.},
        title = "{The Type Ia Supernova Width-Luminosity Relation}",
      journal = {arXiv e-prints},
         year = 2000,
        month = jun,
          eid = {astro-ph/0006171},
        pages = {astro-ph/0006171},
archivePrefix = {arXiv},
       eprint = {astro-ph/0006171},
 primaryClass = {astro-ph},
       adsurl = {https://ui.adsabs.harvard.edu/abs/2000astro.ph..6171P}
}

@ARTICLE{Cowperthwaite17,
       author = {{Cowperthwaite}, P.~S. and {Berger}, E. and {Villar}, V.~A. and
         {Metzger}, B.~D. and {Nicholl}, M. and {Chornock}, R. and
         {Blanchard}, P.~K. and {Fong}, W. and {Margutti}, R. and
         {Soares-Santos}, M. and {Alexander}, K.~D. and {Allam}, S. and
         {Annis}, J. and {Brout}, D. and {Brown}, D.~A. and {Butler}, R.~E. and
         {Chen}, H. -Y. and {Diehl}, H.~T. and {Doctor}, Z. and {Drout}, M.~R. and
         {Eftekhari}, T. and {Farr}, B. and {Finley}, D.~A. and {Foley}, R.~J. and
         {Frieman}, J.~A. and {Fryer}, C.~L. and {Garc{\'\i}a-Bellido}, J. and
         {Gill}, M.~S.~S. and {Guillochon}, J. and {Herner}, K. and
         {Holz}, D.~E. and {Kasen}, D. and {Kessler}, R. and {Marriner}, J. and
         {Matheson}, T. and {Neilsen}, E.~H., Jr. and {Quataert}, E. and
         {Palmese}, A. and {Rest}, A. and {Sako}, M. and {Scolnic}, D.~M. and
         {Smith}, N. and {Tucker}, D.~L. and {Williams}, P.~K.~G. and
         {Balbinot}, E. and {Carlin}, J.~L. and {Cook}, E.~R. and {Durret}, F. and
         {Li}, T.~S. and {Lopes}, P.~A.~A. and {Louren{\c{c}}o}, A.~C.~C. and
         {Marshall}, J.~L. and {Medina}, G.~E. and {Muir}, J. and
         {Mu{\~n}oz}, R.~R. and {Sauseda}, M. and {Schlegel}, D.~J. and
         {Secco}, L.~F. and {Vivas}, A.~K. and {Wester}, W. and {Zenteno}, A. and
         {Zhang}, Y. and {Abbott}, T.~M.~C. and {Banerji}, M. and {Bechtol}, K. and
         {Benoit-L{\'e}vy}, A. and {Bertin}, E. and {Buckley-Geer}, E. and
         {Burke}, D.~L. and {Capozzi}, D. and {Carnero Rosell}, A. and
         {Carrasco Kind}, M. and {Castander}, F.~J. and {Crocce}, M. and
         {Cunha}, C.~E. and {D'Andrea}, C.~B. and {da Costa}, L.~N. and
         {Davis}, C. and {DePoy}, D.~L. and {Desai}, S. and {Dietrich}, J.~P. and
         {Drlica-Wagner}, A. and {Eifler}, T.~F. and {Evrard}, A.~E. and {Fernand
        ez}, E. and {Flaugher}, B. and {Fosalba}, P. and {Gaztanaga}, E. and
         {Gerdes}, D.~W. and {Giannantonio}, T. and {Goldstein}, D.~A. and
         {Gruen}, D. and {Gruendl}, R.~A. and {Gutierrez}, G. and
         {Honscheid}, K. and {Jain}, B. and {James}, D.~J. and {Jeltema}, T. and
         {Johnson}, M.~W.~G. and {Johnson}, M.~D. and {Kent}, S. and
         {Krause}, E. and {Kron}, R. and {Kuehn}, K. and {Nuropatkin}, N. and
         {Lahav}, O. and {Lima}, M. and {Lin}, H. and {Maia}, M.~A.~G. and
         {March}, M. and {Martini}, P. and {McMahon}, R.~G. and {Menanteau}, F. and
         {Miller}, C.~J. and {Miquel}, R. and {Mohr}, J.~J. and {Neilsen}, E. and
         {Nichol}, R.~C. and {Ogando}, R.~L.~C. and {Plazas}, A.~A. and
         {Roe}, N. and {Romer}, A.~K. and {Roodman}, A. and {Rykoff}, E.~S. and
         {Sanchez}, E. and {Scarpine}, V. and {Schindler}, R. and
         {Schubnell}, M. and {Sevilla-Noarbe}, I. and {Smith}, M. and
         {Smith}, R.~C. and {Sobreira}, F. and {Suchyta}, E. and
         {Swanson}, M.~E.~C. and {Tarle}, G. and {Thomas}, D. and
         {Thomas}, R.~C. and {Troxel}, M.~A. and {Vikram}, V. and
         {Walker}, A.~R. and {Wechsler}, R.~H. and {Weller}, J. and {Yanny}, B. and
         {Zuntz}, J.},
        title = "{The Electromagnetic Counterpart of the Binary Neutron Star Merger LIGO/Virgo GW170817. II. UV, Optical, and Near-infrared Light Curves and Comparison to Kilonova Models}",
      journal = {\apjl},
         year = 2017,
        month = oct,
       volume = {848},
       number = {2},
          eid = {L17},
        pages = {L17},
          doi = {10.3847/2041-8213/aa8fc7},
archivePrefix = {arXiv},
       eprint = {1710.05840},
 primaryClass = {astro-ph.HE},
       adsurl = {https://ui.adsabs.harvard.edu/abs/2017ApJ...848L..17C}
}

@ARTICLE{Metzger14,
       author = {{Metzger}, Brian D. and {Fern{\'a}ndez}, Rodrigo},
        title = "{Red or blue? A potential kilonova imprint of the delay until black hole formation following a neutron star merger}",
      journal = {\mnras},
         year = 2014,
        month = jul,
       volume = {441},
       number = {4},
        pages = {3444-3453},
          doi = {10.1093/mnras/stu802},
archivePrefix = {arXiv},
       eprint = {1402.4803},
 primaryClass = {astro-ph.HE},
       adsurl = {https://ui.adsabs.harvard.edu/abs/2014MNRAS.441.3444M}
}

@ARTICLE{Perego14,
       author = {{Perego}, A. and {Rosswog}, S. and {Cabez{\'o}n}, R.~M. and
         {Korobkin}, O. and {K{\"a}ppeli}, R. and {Arcones}, A. and
         {Liebend{\"o}rfer}, M.},
        title = "{Neutrino-driven winds from neutron star merger remnants}",
      journal = {\mnras},
         year = 2014,
        month = oct,
       volume = {443},
       number = {4},
        pages = {3134-3156},
          doi = {10.1093/mnras/stu1352},
archivePrefix = {arXiv},
       eprint = {1405.6730},
 primaryClass = {astro-ph.HE},
       adsurl = {https://ui.adsabs.harvard.edu/abs/2014MNRAS.443.3134P}
}

@ARTICLE{Drout17,
       author = {{Drout}, M.~R. and {Piro}, A.~L. and {Shappee}, B.~J. and
         {Kilpatrick}, C.~D. and {Simon}, J.~D. and {Contreras}, C. and
         {Coulter}, D.~A. and {Foley}, R.~J. and {Siebert}, M.~R. and
         {Morrell}, N. and {Boutsia}, K. and {Di Mille}, F. and
         {Holoien}, T.~W. -S. and {Kasen}, D. and {Kollmeier}, J.~A. and
         {Madore}, B.~F. and {Monson}, A.~J. and {Murguia-Berthier}, A. and
         {Pan}, Y. -C. and {Prochaska}, J.~X. and {Ramirez-Ruiz}, E. and
         {Rest}, A. and {Adams}, C. and {Alatalo}, K. and {Ba{\~n}ados}, E. and
         {Baughman}, J. and {Beers}, T.~C. and {Bernstein}, R.~A. and
         {Bitsakis}, T. and {Campillay}, A. and {Hansen}, T.~T. and
         {Higgs}, C.~R. and {Ji}, A.~P. and {Maravelias}, G. and
         {Marshall}, J.~L. and {Moni Bidin}, C. and {Prieto}, J.~L. and
         {Rasmussen}, K.~C. and {Rojas-Bravo}, C. and {Strom}, A.~L. and
         {Ulloa}, N. and {Vargas-Gonz{\'a}lez}, J. and {Wan}, Z. and
         {Whitten}, D.~D.},
        title = "{Light curves of the neutron star merger GW170817/SSS17a: Implications for r-process nucleosynthesis}",
      journal = {Science},
         year = 2017,
        month = dec,
       volume = {358},
       number = {6370},
        pages = {1570-1574},
          doi = {10.1126/science.aaq0049},
archivePrefix = {arXiv},
       eprint = {1710.05443},
 primaryClass = {astro-ph.HE},
       adsurl = {https://ui.adsabs.harvard.edu/abs/2017Sci...358.1570D}
}

@ARTICLE{Smartt17,
       author = {{Smartt}, S.~J. and {Chen}, T. -W. and {Jerkstrand}, A. and
         {Coughlin}, M. and {Kankare}, E. and {Sim}, S.~A. and {Fraser}, M. and
         {Inserra}, C. and {Maguire}, K. and {Chambers}, K.~C. and
         {Huber}, M.~E. and {Kr{\"u}hler}, T. and {Leloudas}, G. and
         {Magee}, M. and {Shingles}, L.~J. and {Smith}, K.~W. and
         {Young}, D.~R. and {Tonry}, J. and {Kotak}, R. and {Gal-Yam}, A. and
         {Lyman}, J.~D. and {Homan}, D.~S. and {Agliozzo}, C. and
         {Anderson}, J.~P. and {Angus}, C.~R. and {Ashall}, C. and
         {Barbarino}, C. and {Bauer}, F.~E. and {Berton}, M. and
         {Botticella}, M.~T. and {Bulla}, M. and {Bulger}, J. and
         {Cannizzaro}, G. and {Cano}, Z. and {Cartier}, R. and {Cikota}, A. and
         {Clark}, P. and {De Cia}, A. and {Della Valle}, M. and {Denneau}, L. and
         {Dennefeld}, M. and {Dessart}, L. and {Dimitriadis}, G. and
         {Elias-Rosa}, N. and {Firth}, R.~E. and {Flewelling}, H. and
         {Fl{\"o}rs}, A. and {Franckowiak}, A. and {Frohmaier}, C. and
         {Galbany}, L. and {Gonz{\'a}lez-Gait{\'a}n}, S. and {Greiner}, J. and
         {Gromadzki}, M. and {Guelbenzu}, A. Nicuesa and {Guti{\'e}rrez}, C.~P. and
         {Hamanowicz}, A. and {Hanlon}, L. and {Harmanen}, J. and
         {Heintz}, K.~E. and {Heinze}, A. and {Hernandez}, M. -S. and
         {Hodgkin}, S.~T. and {Hook}, I.~M. and {Izzo}, L. and {James}, P.~A. and
         {Jonker}, P.~G. and {Kerzendorf}, W.~E. and {Klose}, S. and
         {Kostrzewa-Rutkowska}, Z. and {Kowalski}, M. and {Kromer}, M. and
         {Kuncarayakti}, H. and {Lawrence}, A. and {Lowe}, T.~B. and
         {Magnier}, E.~A. and {Manulis}, I. and {Martin-Carrillo}, A. and
         {Mattila}, S. and {McBrien}, O. and {M{\"u}ller}, A. and {Nordin}, J. and
         {O'Neill}, D. and {Onori}, F. and {Palmerio}, J.~T. and
         {Pastorello}, A. and {Patat}, F. and {Pignata}, G. and
         {Podsiadlowski}, Ph. and {Pumo}, M.~L. and {Prentice}, S.~J. and
         {Rau}, A. and {Razza}, A. and {Rest}, A. and {Reynolds}, T. and
         {Roy}, R. and {Ruiter}, A.~J. and {Rybicki}, K.~A. and {Salmon}, L. and
         {Schady}, P. and {Schultz}, A.~S.~B. and {Schweyer}, T. and
         {Seitenzahl}, I.~R. and {Smith}, M. and {Sollerman}, J. and
         {Stalder}, B. and {Stubbs}, C.~W. and {Sullivan}, M. and {Szegedi}, H. and
         {Taddia}, F. and {Taubenberger}, S. and {Terreran}, G. and
         {van Soelen}, B. and {Vos}, J. and {Wainscoat}, R.~J. and
         {Walton}, N.~A. and {Waters}, C. and {Weiland}, H. and {Willman}, M. and
         {Wiseman}, P. and {Wright}, D.~E. and {Wyrzykowski}, {\L}. and
         {Yaron}, O.},
        title = "{A kilonova as the electromagnetic counterpart to a gravitational-wave source}",
      journal = {\nat},
         year = 2017,
        month = nov,
       volume = {551},
       number = {7678},
        pages = {75-79},
          doi = {10.1038/nature24303},
archivePrefix = {arXiv},
       eprint = {1710.05841},
 primaryClass = {astro-ph.HE},
       adsurl = {https://ui.adsabs.harvard.edu/abs/2017Natur.551...75S}
}

@ARTICLE{Coulter17,
       author = {{Coulter}, D.~A. and {Foley}, R.~J. and {Kilpatrick}, C.~D. and
         {Drout}, M.~R. and {Piro}, A.~L. and {Shappee}, B.~J. and
         {Siebert}, M.~R. and {Simon}, J.~D. and {Ulloa}, N. and {Kasen}, D. and
         {Madore}, B.~F. and {Murguia-Berthier}, A. and {Pan}, Y. -C. and
         {Prochaska}, J.~X. and {Ramirez-Ruiz}, E. and {Rest}, A. and
         {Rojas-Bravo}, C.},
        title = "{Swope Supernova Survey 2017a (SSS17a), the optical counterpart to a gravitational wave source}",
      journal = {Science},
         year = 2017,
        month = dec,
       volume = {358},
       number = {6370},
        pages = {1556-1558},
          doi = {10.1126/science.aap9811},
archivePrefix = {arXiv},
       eprint = {1710.05452},
 primaryClass = {astro-ph.HE},
       adsurl = {https://ui.adsabs.harvard.edu/abs/17Sci...358.1556C}
}

@ARTICLE{Just15,
       author = {{Just}, O. and {Bauswein}, A. and {Ardevol Pulpillo}, R. and
         {Goriely}, S. and {Janka}, H. -T.},
        title = "{Comprehensive nucleosynthesis analysis for ejecta of compact binary mergers}",
      journal = {\mnras},
         year = 2015,
        month = mar,
       volume = {448},
       number = {1},
        pages = {541-567},
          doi = {10.1093/mnras/stv009},
archivePrefix = {arXiv},
       eprint = {1406.2687},
 primaryClass = {astro-ph.SR},
       adsurl = {https://ui.adsabs.harvard.edu/abs/2015MNRAS.448..541J}
}

@ARTICLE{Just22,
       author = {{Just}, O. and {Kullmann}, I. and {Goriely}, S. and {Bauswein}, A. and {Janka}, H. -T. and {Collins}, C.~E.},
        title = "{Dynamical ejecta of neutron star mergers with nucleonic weak processes - II: kilonova emission}",
      journal = {\mnras},
         year = 2022,
        month = feb,
       volume = {510},
       number = {2},
        pages = {2820-2840},
          doi = {10.1093/mnras/stab3327},
archivePrefix = {arXiv},
       eprint = {2109.14617},
 primaryClass = {astro-ph.HE},
       adsurl = {https://ui.adsabs.harvard.edu/abs/2022MNRAS.510.2820J}
}

@ARTICLE{Lattimer74,
       author = {{Lattimer}, J.~M. and {Schramm}, D.~N.},
        title = "{Black-Hole-Neutron-Star Collisions}",
      journal = {\apjl},
         year = 1974,
        month = sep,
       volume = {192},
        pages = {L145},
          doi = {10.1086/181612},
       adsurl = {https://ui.adsabs.harvard.edu/abs/1974ApJ...192L.145L}
}

@ARTICLE{Eichler89,
       author = {{Eichler}, David and {Livio}, Mario and {Piran}, Tsvi and
         {Schramm}, David N.},
        title = "{Nucleosynthesis, neutrino bursts and {\ensuremath{\gamma}}-rays from coalescing neutron stars}",
      journal = {\nat},
         year = 1989,
        month = jul,
       volume = {340},
       number = {6229},
        pages = {126-128},
          doi = {10.1038/340126a0},
       adsurl = {https://ui.adsabs.harvard.edu/abs/1989Natur.340..126E}
}

@ARTICLE{Freiburghaus99,
       author = {{Freiburghaus}, C. and {Rosswog}, S. and {Thielemann}, F. -K.},
        title = "{R-Process in Neutron Star Mergers}",
      journal = {\apjl},
         year = 1999,
        month = nov,
       volume = {525},
       number = {2},
        pages = {L121-L124},
          doi = {10.1086/312343},
       adsurl = {https://ui.adsabs.harvard.edu/abs/1999ApJ...525L.121F}
}

@INPROCEEDINGS{Savchenko17,
       author = {{Savchenko}, V. and {Ferrigno}, C. and {Kuulkers}, E. and {Bazzano}, A. and
         {Bozzo}, E. and {Brandt}, S. and {Chenevez}, J. and {Diehl}, R. and
         {Domingo}, A. and {Hanlon}, L. and {Jourdain}, E. and {Laurent}, P. and
         {Lebrun}, F. and {Lutovinov}, A. and {Martin-Carillo}, A. and
         {Mereghetti}, S. and {Natalucci}, L. and {Rodi}, J. and {Sunyaev}, R. and
         {Ubertini}, P.},
        title = "{INTEGRAL follow-up of the gravitational wave events}",
    booktitle = {Proceedings of the 7th International Fermi Symposium},
         year = 2017,
        month = oct,
          eid = {58},
        pages = {58},
       adsurl = {https://ui.adsabs.harvard.edu/abs/2017ifs..confE..58S}
}

@INPROCEEDINGS{Connaughton17,
       author = {{Connaughton}, Valerie and {Goldstein}, Adam and
         {Fermi GBM - LIGO Group}},
        title = "{Multi-Messenger Time-Domain Astronomy with the Fermi Gamma-ray Burst Monitor}",
    booktitle = {American Astronomical Society Meeting Abstracts \#229},
         year = 2017,
       series = {American Astronomical Society Meeting Abstracts},
       volume = {229},
        month = jan,
          eid = {406.08},
        pages = {406.08},
       adsurl = {https://ui.adsabs.harvard.edu/abs/2017AAS...22940608C}
}

@ARTICLE{Valenti17,
       author = {{Valenti}, Stefano and {Sand}, David J. and {Yang}, Sheng and
         {Cappellaro}, Enrico and {Tartaglia}, Leonardo and {Corsi}, Alessandra and
         {Jha}, Saurabh W. and {Reichart}, Daniel E. and {Haislip}, Joshua and
         {Kouprianov}, Vladimir},
        title = "{The Discovery of the Electromagnetic Counterpart of GW170817: Kilonova AT 2017gfo/DLT17ck}",
      journal = {\apjl},
         year = 2017,
        month = oct,
       volume = {848},
       number = {2},
          eid = {L24},
        pages = {L24},
          doi = {10.3847/2041-8213/aa8edf},
archivePrefix = {arXiv},
       eprint = {1710.05854},
 primaryClass = {astro-ph.HE},
       adsurl = {https://ui.adsabs.harvard.edu/abs/2017ApJ...848L..24V}
}

@ARTICLE{Yang17,
       author = {{Yang}, Sheng and {Valenti}, Stefano and {Cappellaro}, Enrico and {Sand
        }, David J. and {Tartaglia}, Leonardo and {Corsi}, Alessandra and
         {Reichart}, Daniel E. and {Haislip}, Joshua and {Kouprianov}, Vladimir},
        title = "{An Empirical Limit on the Kilonova Rate from the DLT40 One Day Cadence Supernova Survey}",
      journal = {\apjl},
         year = 2017,
        month = dec,
       volume = {851},
       number = {2},
          eid = {L48},
        pages = {L48},
          doi = {10.3847/2041-8213/aaa07d},
archivePrefix = {arXiv},
       eprint = {1710.05864},
 primaryClass = {astro-ph.HE},
       adsurl = {https://ui.adsabs.harvard.edu/abs/2017ApJ...851L..48Y}
}

@ARTICLE{Korobkin12,
       author = {{Korobkin}, O. and {Rosswog}, S. and {Arcones}, A. and {Winteler}, C.},
        title = "{On the astrophysical robustness of the neutron star merger r-process}",
      journal = {\mnras},
         year = 2012,
        month = nov,
       volume = {426},
       number = {3},
        pages = {1940-1949},
          doi = {10.1111/j.1365-2966.2012.21859.x},
archivePrefix = {arXiv},
       eprint = {1206.2379},
 primaryClass = {astro-ph.SR},
       adsurl = {https://ui.adsabs.harvard.edu/abs/2012MNRAS.426.1940K}
}

@ARTICLE{Fujibayashi18,
       author = {{Fujibayashi}, Sho and {Kiuchi}, Kenta and {Nishimura}, Nobuya and
         {Sekiguchi}, Yuichiro and {Shibata}, Masaru},
        title = "{Mass Ejection from the Remnant of a Binary Neutron Star Merger: Viscous-radiation Hydrodynamics Study}",
      journal = {\apj},
         year = 2018,
        month = jun,
       volume = {860},
       number = {1},
          eid = {64},
        pages = {64},
          doi = {10.3847/1538-4357/aabafd},
archivePrefix = {arXiv},
       eprint = {1711.02093},
 primaryClass = {astro-ph.HE},
       adsurl = {https://ui.adsabs.harvard.edu/abs/2018ApJ...860...64F}
}

@ARTICLE{Lippuner17,
       author = {{Lippuner}, Jonas and {Fern{\'a}ndez}, Rodrigo and {Roberts}, Luke F. and
         {Foucart}, Francois and {Kasen}, Daniel and {Metzger}, Brian D. and
         {Ott}, Christian D.},
        title = "{Signatures of hypermassive neutron star lifetimes on r-process nucleosynthesis in the disc ejecta from neutron star mergers}",
      journal = {\mnras},
         year = 2017,
        month = nov,
       volume = {472},
       number = {1},
        pages = {904-918},
          doi = {10.1093/mnras/stx1987},
archivePrefix = {arXiv},
       eprint = {1703.06216},
 primaryClass = {astro-ph.HE},
       adsurl = {https://ui.adsabs.harvard.edu/abs/2017MNRAS.472..904L}
}

@ARTICLE{Fernandez13,
       author = {{Fern{\'a}ndez}, Rodrigo and {Metzger}, Brian D.},
        title = "{Delayed outflows from black hole accretion tori following neutron star binary coalescence}",
      journal = {\mnras},
         year = 2013,
        month = oct,
       volume = {435},
       number = {1},
        pages = {502-517},
          doi = {10.1093/mnras/stt1312},
archivePrefix = {arXiv},
       eprint = {1304.6720},
 primaryClass = {astro-ph.HE},
       adsurl = {https://ui.adsabs.harvard.edu/abs/2013MNRAS.435..502F}
}

@ARTICLE{Sagiv14,
       author = {{Sagiv}, I. and {Gal-Yam}, A. and {Ofek}, E.~O. and {Waxman}, E. and
         {Aharonson}, O. and {Kulkarni}, S.~R. and {Nakar}, E. and {Maoz}, D. and
         {Trakhtenbrot}, B. and {Phinney}, E.~S. and {Topaz}, J. and
         {Beichman}, C. and {Murthy}, J. and {Worden}, S.~P.},
        title = "{Science with a Wide-field UV Transient Explorer}",
      journal = {\aj},
         year = 2014,
        month = apr,
       volume = {147},
       number = {4},
          eid = {79},
        pages = {79},
          doi = {10.1088/0004-6256/147/4/79},
archivePrefix = {arXiv},
       eprint = {1303.6194},
 primaryClass = {astro-ph.CO},
       adsurl = {https://ui.adsabs.harvard.edu/abs/2014AJ....147...79S}
}

@ARTICLE{Smaranikab20,
       author = {{Banerjee}, Smaranika and {Tanaka}, Masaomi and {Kawaguchi}, Kyohei and {Kato}, Daiji and {Gaigalas}, Gediminas},
        title = "{Simulations of Early Kilonova Emission from Neutron Star Mergers}",
      journal = {\apj},
         year = 2020,
        month = sep,
       volume = {901},
       number = {1},
          eid = {29},
        pages = {29},
          doi = {10.3847/1538-4357/abae61},
archivePrefix = {arXiv},
       eprint = {2008.05495},
 primaryClass = {astro-ph.HE},
       adsurl = {https://ui.adsabs.harvard.edu/abs/2020ApJ...901...29B}
}

@ARTICLE{Sobolev60,
       author = {{Sobolev}, V.~V.},
        title = "{The Theory of Stellar Evolution}",
      journal = {\sovast},
         year = 1960,
        month = dec,
       volume = {4},
        pages = {372},
       adsurl = {https://ui.adsabs.harvard.edu/abs/1960SvA.....4..372S}
}

@ARTICLE{Duffell18,
       author = {{Duffell}, Paul C. and {Quataert}, Eliot and {Kasen}, Daniel and {Klion}, Hannah},
        title = "{Jet Dynamics in Compact Object Mergers: GW170817 Likely Had a Successful Jet}",
      journal = {\apj},
         year = 2018,
        month = oct,
       volume = {866},
       number = {1},
          eid = {3},
        pages = {3},
          doi = {10.3847/1538-4357/aae084},
archivePrefix = {arXiv},
       eprint = {1806.10616},
 primaryClass = {astro-ph.HE},
       adsurl = {https://ui.adsabs.harvard.edu/abs/2018ApJ...866....3D}
}

@ARTICLE{Klion21,
       author = {{Klion}, Hannah and {Duffell}, Paul C. and {Kasen}, Daniel and {Quataert}, Eliot},
        title = "{The effect of jet-ejecta interaction on the viewing angle dependence of kilonova light curves}",
      journal = {\mnras},
         year = 2021,
        month = mar,
       volume = {502},
       number = {1},
        pages = {865-875},
          doi = {10.1093/mnras/stab042},
archivePrefix = {arXiv},
       eprint = {2012.08577},
 primaryClass = {astro-ph.HE},
       adsurl = {https://ui.adsabs.harvard.edu/abs/2021MNRAS.502..865K}
}

@ARTICLE{Nativi21,
       author = {{Nativi}, L. and {Bulla}, M. and {Rosswog}, S. and {Lundman}, C. and {Kowal}, G. and {Gizzi}, D. and {Lamb}, G.~P. and {Perego}, A.},
        title = "{Can jets make the radioactively powered emission from neutron star mergers bluer?}",
      journal = {\mnras},
         year = 2021,
        month = jan,
       volume = {500},
       number = {2},
        pages = {1772-1783},
          doi = {10.1093/mnras/staa3337},
archivePrefix = {arXiv},
       eprint = {2010.08989},
 primaryClass = {astro-ph.HE},
       adsurl = {https://ui.adsabs.harvard.edu/abs/2021MNRAS.500.1772N}
}

@ARTICLE{Kiuchi19,
       author = {{Kiuchi}, Kenta and {Kyutoku}, Koutarou and {Shibata}, Masaru and {Taniguchi}, Keisuke},
        title = "{Revisiting the Lower Bound on Tidal Deformability Derived by AT 2017gfo}",
      journal = {\apjl},
         year = 2019,
        month = may,
       volume = {876},
       number = {2},
          eid = {L31},
        pages = {L31},
          doi = {10.3847/2041-8213/ab1e45},
archivePrefix = {arXiv},
       eprint = {1903.01466},
 primaryClass = {astro-ph.HE},
       adsurl = {https://ui.adsabs.harvard.edu/abs/2019ApJ...876L..31K}
}

@ARTICLE{Goldstein17,
       author = {{Goldstein}, A. and {Veres}, P. and {Burns}, E. and {Briggs}, M.~S. and {Hamburg}, R. and {Kocevski}, D. and {Wilson-Hodge}, C.~A. and {Preece}, R.~D. and {Poolakkil}, S. and {Roberts}, O.~J. and {Hui}, C.~M. and {Connaughton}, V. and {Racusin}, J. and {von Kienlin}, A. and {Dal Canton}, T. and {Christensen}, N. and {Littenberg}, T. and {Siellez}, K. and {Blackburn}, L. and {Broida}, J. and {Bissaldi}, E. and {Cleveland}, W.~H. and {Gibby}, M.~H. and {Giles}, M.~M. and {Kippen}, R.~M. and {McBreen}, S. and {McEnery}, J. and {Meegan}, C.~A. and {Paciesas}, W.~S. and {Stanbro}, M.},
        title = "{An Ordinary Short Gamma-Ray Burst with Extraordinary Implications: Fermi-GBM Detection of GRB 170817A}",
      journal = {\apjl},
         year = 2017,
        month = oct,
       volume = {848},
       number = {2},
          eid = {L14},
        pages = {L14},
          doi = {10.3847/2041-8213/aa8f41},
archivePrefix = {arXiv},
       eprint = {1710.05446},
 primaryClass = {astro-ph.HE},
       adsurl = {https://ui.adsabs.harvard.edu/abs/2017ApJ...848L..14G}
}

@ARTICLE{Escorial23,
       author = {{Rouco Escorial}, A. and {Fong}, W. and {Berger}, E. and {Laskar}, T. and {Margutti}, R. and {Schroeder}, G. and {Rastinejad}, J.~C. and {Cornish}, D. and {Popp}, S. and {Lally}, M. and {Nugent}, A.~E. and {Paterson}, K. and {Metzger}, B.~D. and {Chornock}, R. and {Alexander}, K. and {Cendes}, Y. and {Eftekhari}, T.},
        title = "{The Jet Opening Angle and Event Rate Distributions of Short Gamma-Ray Bursts from Late-time X-Ray Afterglows}",
      journal = {\apj},
         year = 2023,
        month = dec,
       volume = {959},
       number = {1},
          eid = {13},
        pages = {13},
          doi = {10.3847/1538-4357/acf830},
archivePrefix = {arXiv},
       eprint = {2210.05695},
 primaryClass = {astro-ph.HE},
       adsurl = {https://ui.adsabs.harvard.edu/abs/2023ApJ...959...13R}
}

@ARTICLE{Kawaguchi21,
       author = {{Kawaguchi}, Kyohei and {Fujibayashi}, Sho and {Shibata}, Masaru and {Tanaka}, Masaomi and {Wanajo}, Shinya},
        title = "{A Low-mass Binary Neutron Star: Long-term Ejecta Evolution and Kilonovae with Weak Blue Emission}",
      journal = {\apj},
         year = 2021,
        month = jun,
       volume = {913},
       number = {2},
          eid = {100},
        pages = {100},
          doi = {10.3847/1538-4357/abf3bc},
archivePrefix = {arXiv},
       eprint = {2012.14711},
 primaryClass = {astro-ph.HE},
       adsurl = {https://ui.adsabs.harvard.edu/abs/2021ApJ...913..100K}
}

@ARTICLE{Kiuchi23,
       author = {{Kiuchi}, Kenta and {Fujibayashi}, Sho and {Hayashi}, Kota and {Kyutoku}, Koutarou and {Sekiguchi}, Yuichiro and {Shibata}, Masaru},
        title = "{Self-Consistent Picture of the Mass Ejection from a One Second Long Binary Neutron Star Merger Leaving a Short-Lived Remnant in a General-Relativistic Neutrino-Radiation Magnetohydrodynamic Simulation}",
      journal = {\prl},
         year = 2023,
        month = jul,
       volume = {131},
       number = {1},
          eid = {011401},
        pages = {011401},
          doi = {10.1103/PhysRevLett.131.011401},
archivePrefix = {arXiv},
       eprint = {2211.07637},
 primaryClass = {astro-ph.HE},
       adsurl = {https://ui.adsabs.harvard.edu/abs/2023PhRvL.131a1401K}
}

@ARTICLE{Bulla19,
       author = {{Bulla}, M.},
        title = "{POSSIS: predicting spectra, light curves, and polarization for multidimensional models of supernovae and kilonovae}",
      journal = {\mnras},
         year = 2019,
        month = nov,
       volume = {489},
       number = {4},
        pages = {5037-5045},
          doi = {10.1093/mnras/stz2495},
archivePrefix = {arXiv},
       eprint = {1906.04205},
 primaryClass = {astro-ph.HE},
       adsurl = {https://ui.adsabs.harvard.edu/abs/2019MNRAS.489.5037B}
}

@ARTICLE{Shrestha23,
       author = {{Shrestha}, Manisha and {Bulla}, Mattia and {Nativi}, Lorenzo and {Markin}, Ivan and {Rosswog}, Stephan and {Dietrich}, Tim},
        title = "{Impact of jets on kilonova photometric and polarimetric emission from binary neutron star mergers}",
      journal = {\mnras},
         year = 2023,
        month = aug,
       volume = {523},
       number = {2},
        pages = {2990-3000},
          doi = {10.1093/mnras/stad1583},
archivePrefix = {arXiv},
       eprint = {2303.14277},
 primaryClass = {astro-ph.HE},
       adsurl = {https://ui.adsabs.harvard.edu/abs/2023MNRAS.523.2990S}
}

@ARTICLE{Wanajo18,
       author = {{Wanajo}, Shinya and {M{\"u}ller}, Bernhard and {Janka}, Hans-Thomas and {Heger}, Alexander},
        title = "{Nucleosynthesis in the Innermost Ejecta of Neutrino-driven Supernova Explosions in Two Dimensions}",
      journal = {\apj},
         year = 2018,
        month = jan,
       volume = {852},
       number = {1},
          eid = {40},
        pages = {40},
          doi = {10.3847/1538-4357/aa9d97},
archivePrefix = {arXiv},
       eprint = {1701.06786},
 primaryClass = {astro-ph.SR},
       adsurl = {https://ui.adsabs.harvard.edu/abs/2018ApJ...852...40W}
}

@ARTICLE{Smaranikab25,
       author = {{Banerjee}, Smaranika and {Jerkstrand}, Anders and {Badnell}, Nigel and {Pognan}, Quentin and {Ferguson}, Niamh and {Grumor}, Jon},
        title = "{Nebular spectra of kilonovae with detailed recombination rates -- I. Light r-process composition}",
      journal = {arXiv e-prints},
         year = 2025,
        month = jan,
          eid = {arXiv:2501.18345},
        pages = {arXiv:2501.18345},
          doi = {10.48550/arXiv.2501.18345},
archivePrefix = {arXiv},
       eprint = {2501.18345},
 primaryClass = {astro-ph.HE},
       adsurl = {https://ui.adsabs.harvard.edu/abs/2025arXiv250118345B}
}

@ARTICLE{Kasen06,
       author = {{Kasen}, Daniel and {Thomas}, R.~C. and {Nugent}, P.},
        title = "{Time-dependent Monte Carlo Radiative Transfer Calculations for Three-dimensional Supernova Spectra, Light Curves, and Polarization}",
      journal = {\apj},
         year = 2006,
        month = nov,
       volume = {651},
       number = {1},
        pages = {366-380},
          doi = {10.1086/506190},
archivePrefix = {arXiv},
       eprint = {astro-ph/0606111},
 primaryClass = {astro-ph},
       adsurl = {https://ui.adsabs.harvard.edu/abs/2006ApJ...651..366K}
}

@ARTICLE{Hamidani23a,
       author = {{Hamidani}, Hamid and {Ioka}, Kunihito},
        title = "{Cocoon breakout and escape from the ejecta of neutron star mergers}",
      journal = {\mnras},
         year = 2023,
        month = mar,
       volume = {520},
       number = {1},
        pages = {1111-1127},
          doi = {10.1093/mnras/stad041},
archivePrefix = {arXiv},
       eprint = {2210.00814},
 primaryClass = {astro-ph.HE},
       adsurl = {https://ui.adsabs.harvard.edu/abs/2023MNRAS.520.1111H}
}

@ARTICLE{Hamidani23b,
       author = {{Hamidani}, Hamid and {Ioka}, Kunihito},
        title = "{Cocoon cooling emission in neutron star mergers}",
      journal = {\mnras},
         year = 2023,
        month = oct,
       volume = {524},
       number = {4},
        pages = {4841-4866},
          doi = {10.1093/mnras/stad1933},
archivePrefix = {arXiv},
       eprint = {2210.02255},
 primaryClass = {astro-ph.HE},
       adsurl = {https://ui.adsabs.harvard.edu/abs/2023MNRAS.524.4841H}
}

@ARTICLE{Fong15,
       author = {{Fong}, W. and {Berger}, E. and {Margutti}, R. and {Zauderer}, B.~A.},
        title = "{A Decade of Short-duration Gamma-Ray Burst Broadband Afterglows: Energetics, Circumburst Densities, and Jet Opening Angles}",
      journal = {\apj},
         year = 2015,
        month = dec,
       volume = {815},
       number = {2},
          eid = {102},
        pages = {102},
          doi = {10.1088/0004-637X/815/2/102},
archivePrefix = {arXiv},
       eprint = {1509.02922},
 primaryClass = {astro-ph.HE},
       adsurl = {https://ui.adsabs.harvard.edu/abs/2015ApJ...815..102F}
}

@ARTICLE{Hamidani20,
       author = {{Hamidani}, Hamid and {Kiuchi}, Kenta and {Ioka}, Kunihito},
        title = "{Jet propagation in neutron star mergers and GW170817}",
      journal = {\mnras},
         year = 2020,
        month = jan,
       volume = {491},
       number = {3},
        pages = {3192-3216},
          doi = {10.1093/mnras/stz3231},
archivePrefix = {arXiv},
       eprint = {1909.05867},
 primaryClass = {astro-ph.HE},
       adsurl = {https://ui.adsabs.harvard.edu/abs/2020MNRAS.491.3192H}
}

@ARTICLE{Ryan20,
       author = {{Ryan}, Geoffrey and {van Eerten}, Hendrik and {Piro}, Luigi and {Troja}, Eleonora},
        title = "{Gamma-Ray Burst Afterglows in the Multimessenger Era: Numerical Models and Closure Relations}",
      journal = {\apj},
         year = 2020,
        month = jun,
       volume = {896},
       number = {2},
          eid = {166},
        pages = {166},
          doi = {10.3847/1538-4357/ab93cf},
archivePrefix = {arXiv},
       eprint = {1909.11691},
 primaryClass = {astro-ph.HE},
       adsurl = {https://ui.adsabs.harvard.edu/abs/2020ApJ...896..166R}
}

@ARTICLE{Kulkarni21,
       author = {{Kulkarni}, S.~R. and {Harrison}, Fiona A. and {Grefenstette}, Brian W. and {Earnshaw}, Hannah P. and {Andreoni}, Igor and {Berg}, Danielle A. and {Bloom}, Joshua S. and {Cenko}, S. Bradley and {Chornock}, Ryan and {Christiansen}, Jessie L. and {Coughlin}, Michael W. and {Wuollet Criswell}, Alexander and {Darvish}, Behnam and {Das}, Kaustav K. and {De}, Kishalay and {Dessart}, Luc and {Dixon}, Don and {Dorsman}, Bas and {El-Badry}, Kareem and {Evans}, Christopher and {Ford}, K.~E. Saavik and {Fremling}, Christoffer and {Gansicke}, Boris T. and {Gezari}, Suvi and {Gotberg}, Y. and {Green}, Gregory M. and {Graham}, Matthew J. and {Heida}, Marianne and {Ho}, Anna Y.~Q. and {Jaodand}, Amruta D. and {Johns-Krull}, Christopher M. and {Kasliwal}, Mansi M. and {Lazzarini}, Margaret and {Lu}, Wenbin and {Margutti}, Raffaella and {Martin}, D. Christopher and {Masters}, Daniel Charles and {McKernan}, Barry and {Nissanke}, Samaya M. and {Parazin}, B. and {Perley}, Daniel A. and {Phinney}, E. Sterl and {Piro}, Anthony L. and {Raaijmakers}, G. and {Rodriguez}, Antonio C. and {Senchyna}, Peter and {Singer}, Leo P. and {Spake}, Jessica J. and {Stassun}, Keivan G. and {Stern}, Daniel and {Teplitz}, Harry I. and {Weisz}, Daniel R. and {Yao}, Yuhan},
        title = "{Science with the Ultraviolet Explorer (UVEX)}",
      journal = {arXiv e-prints},
         year = 2021,
        month = nov,
          eid = {arXiv:2111.15608},
        pages = {arXiv:2111.15608},
archivePrefix = {arXiv},
       eprint = {2111.15608},
 primaryClass = {astro-ph.GA},
       adsurl = {https://ui.adsabs.harvard.edu/abs/2021arXiv211115608K}
}

@ARTICLE{Kromer09,
       author = {{Kromer}, M. and {Sim}, S.~A.},
        title = "{Time-dependent three-dimensional spectrum synthesis for Type Ia supernovae}",
      journal = {\mnras},
         year = 2009,
        month = oct,
       volume = {398},
       number = {4},
        pages = {1809-1826},
          doi = {10.1111/j.1365-2966.2009.15256.x},
archivePrefix = {arXiv},
       eprint = {0906.3152},
 primaryClass = {astro-ph.HE},
       adsurl = {https://ui.adsabs.harvard.edu/abs/2009MNRAS.398.1809K}
}

@ARTICLE{Mooley18,
       author = {{Mooley}, K.~P. and {Deller}, A.~T. and {Gottlieb}, O. and {Nakar}, E. and {Hallinan}, G. and {Bourke}, S. and {Frail}, D.~A. and {Horesh}, A. and {Corsi}, A. and {Hotokezaka}, K.},
        title = "{Superluminal motion of a relativistic jet in the neutron-star merger GW170817}",
      journal = {\nat},
         year = 2018,
        month = sep,
       volume = {561},
       number = {7723},
        pages = {355-359},
          doi = {10.1038/s41586-018-0486-3},
archivePrefix = {arXiv},
       eprint = {1806.09693},
 primaryClass = {astro-ph.HE},
       adsurl = {https://ui.adsabs.harvard.edu/abs/2018Natur.561..355M}
}

@ARTICLE{Finstad18,
       author = {{Finstad}, Daniel and {De}, Soumi and {Brown}, Duncan A. and {Berger}, Edo and {Biwer}, Christopher M.},
        title = "{Measuring the Viewing Angle of GW170817 with Electromagnetic and Gravitational Waves}",
      journal = {\apjl},
         year = 2018,
        month = jun,
       volume = {860},
       number = {1},
          eid = {L2},
        pages = {L2},
          doi = {10.3847/2041-8213/aac6c1},
archivePrefix = {arXiv},
       eprint = {1804.04179},
 primaryClass = {astro-ph.HE},
       adsurl = {https://ui.adsabs.harvard.edu/abs/2018ApJ...860L...2F}
}

@ARTICLE{Rastinejad22,
       author = {{Rastinejad}, J.~C. and {Gompertz}, B.~P. and {Levan}, A.~J. and {Fong}, W. and {Nicholl}, M. and {Lamb}, G.~P. and {Malesani}, D.~B. and {Nugent}, A.~E. and {Oates}, S.~R. and {Tanvir}, N.~R. and {de Ugarte Postigo}, A. and {Kilpatrick}, C.~D. and {Moore}, C.~J. and {Metzger}, B.~D. and {Ravasio}, M.~E. and {Rossi}, A. and {Schroeder}, G. and {Jencson}, J. and {Sand}, D.~J. and {Smith}, N. and {Ag{\"u}{\'\i} Fern{\'a}ndez}, J.~F. and {Berger}, E. and {Blanchard}, P.~K. and {Chornock}, R. and {Cobb}, B.~E. and {De Pasquale}, M. and {Fynbo}, J.~P.~U. and {Izzo}, L. and {Kann}, D.~A. and {Laskar}, T. and {Marini}, E. and {Paterson}, K. and {Rouco Escorial}, A. and {Sears}, H.~M. and {Th{\"o}ne}, C.~C.},
        title = "{A Kilonova Following a Long-Duration Gamma-Ray Burst at 350 Mpc}",
      journal = {arXiv e-prints},
         year = 2022,
        month = apr,
          eid = {arXiv:2204.10864},
        pages = {arXiv:2204.10864},
archivePrefix = {arXiv},
       eprint = {2204.10864},
 primaryClass = {astro-ph.HE},
       adsurl = {https://ui.adsabs.harvard.edu/abs/2022arXiv220410864R}
}

@ARTICLE{Miller19,
       author = {{Miller}, Jonah M. and {Ryan}, Benjamin R. and {Dolence}, Joshua C. and {Burrows}, Adam and {Fontes}, Christopher J. and {Fryer}, Christopher L. and {Korobkin}, Oleg and {Lippuner}, Jonas and {Mumpower}, Matthew R. and {Wollaeger}, Ryan T.},
        title = "{Full transport model of GW170817-like disk produces a blue kilonova}",
      journal = {\prd},
         year = 2019,
        month = jul,
       volume = {100},
       number = {2},
          eid = {023008},
        pages = {023008},
          doi = {10.1103/PhysRevD.100.023008},
archivePrefix = {arXiv},
       eprint = {1905.07477},
 primaryClass = {astro-ph.HE},
       adsurl = {https://ui.adsabs.harvard.edu/abs/2019PhRvD.100b3008M}
}

@ARTICLE{Metzger20,
       author = {{Metzger}, Brian D.},
        title = "{Kilonovae}",
      journal = {Living Reviews in Relativity},
         year = 2020,
        month = dec,
       volume = {23},
       number = {1},
          eid = {1},
        pages = {1},
          doi = {10.1007/s41114-019-0024-0},
       adsurl = {https://ui.adsabs.harvard.edu/abs/2020LRR....23....1M}
}

@ARTICLE{Lazzati20,
       author = {{Lazzati}, Davide},
        title = "{Short Duration Gamma-Ray Bursts and Their Outflows in Light of GW170817}",
      journal = {Frontiers in Astronomy and Space Sciences},
         year = 2020,
        month = nov,
       volume = {7},
          eid = {78},
        pages = {78},
          doi = {10.3389/fspas.2020.578849},
archivePrefix = {arXiv},
       eprint = {2009.01773},
 primaryClass = {astro-ph.HE},
       adsurl = {https://ui.adsabs.harvard.edu/abs/2020FrASS...7...78L}
}

@ARTICLE{Tanvir17,
       author = {{Tanvir}, N.~R. and {Levan}, A.~J. and {Gonz{\'a}lez-Fern{\'a}ndez}, C. and {Korobkin}, O. and {Mandel}, I. and {Rosswog}, S. and {Hjorth}, J. and {D'Avanzo}, P. and {Fruchter}, A.~S. and {Fryer}, C.~L. and {Kangas}, T. and {Milvang-Jensen}, B. and {Rosetti}, S. and {Steeghs}, D. and {Wollaeger}, R.~T. and {Cano}, Z. and {Copperwheat}, C.~M. and {Covino}, S. and {D'Elia}, V. and {de Ugarte Postigo}, A. and {Evans}, P.~A. and {Even}, W.~P. and {Fairhurst}, S. and {Figuera Jaimes}, R. and {Fontes}, C.~J. and {Fujii}, Y.~I. and {Fynbo}, J.~P.~U. and {Gompertz}, B.~P. and {Greiner}, J. and {Hodosan}, G. and {Irwin}, M.~J. and {Jakobsson}, P. and {J{\o}rgensen}, U.~G. and {Kann}, D.~A. and {Lyman}, J.~D. and {Malesani}, D. and {McMahon}, R.~G. and {Melandri}, A. and {O'Brien}, P.~T. and {Osborne}, J.~P. and {Palazzi}, E. and {Perley}, D.~A. and {Pian}, E. and {Piranomonte}, S. and {Rabus}, M. and {Rol}, E. and {Rowlinson}, A. and {Schulze}, S. and {Sutton}, P. and {Th{\"o}ne}, C.~C. and {Ulaczyk}, K. and {Watson}, D. and {Wiersema}, K. and {Wijers}, R.~A.~M.~J.},
        title = "{The Emergence of a Lanthanide-rich Kilonova Following the Merger of Two Neutron Stars}",
      journal = {\apjl},
         year = 2017,
        month = oct,
       volume = {848},
       number = {2},
          eid = {L27},
        pages = {L27},
          doi = {10.3847/2041-8213/aa90b6},
archivePrefix = {arXiv},
       eprint = {1710.05455},
 primaryClass = {astro-ph.HE},
       adsurl = {https://ui.adsabs.harvard.edu/abs/2017ApJ...848L..27T}
}

@ARTICLE{Levan24,
       author = {{Levan}, Andrew J. and {Gompertz}, Benjamin P. and {Salafia}, Om Sharan and {Bulla}, Mattia and {Burns}, Eric and {Hotokezaka}, Kenta and {Izzo}, Luca and {Lamb}, Gavin P. and {Malesani}, Daniele B. and {Oates}, Samantha R. and {Ravasio}, Maria Edvige and {Rouco Escorial}, Alicia and {Schneider}, Benjamin and {Sarin}, Nikhil and {Schulze}, Steve and {Tanvir}, Nial R. and {Ackley}, Kendall and {Anderson}, Gemma and {Brammer}, Gabriel B. and {Christensen}, Lise and {Dhillon}, Vikram S. and {Evans}, Phil A. and {Fausnaugh}, Michael and {Fong}, Wen-fai and {Fruchter}, Andrew S. and {Fryer}, Chris and {Fynbo}, Johan P.~U. and {Gaspari}, Nicola and {Heintz}, Kasper E. and {Hjorth}, Jens and {Kennea}, Jamie A. and {Kennedy}, Mark R. and {Laskar}, Tanmoy and {Leloudas}, Giorgos and {Mandel}, Ilya and {Martin-Carrillo}, Antonio and {Metzger}, Brian D. and {Nicholl}, Matt and {Nugent}, Anya and {Palmerio}, Jesse T. and {Pugliese}, Giovanna and {Rastinejad}, Jillian and {Rhodes}, Lauren and {Rossi}, Andrea and {Saccardi}, Andrea and {Smartt}, Stephen J. and {Stevance}, Heloise F. and {Tohuvavohu}, Aaron and {van der Horst}, Alexander and {Vergani}, Susanna D. and {Watson}, Darach and {Barclay}, Thomas and {Bhirombhakdi}, Kornpob and {Breedt}, Elm{\'e} and {Breeveld}, Alice A. and {Brown}, Alexander J. and {Campana}, Sergio and {Chrimes}, Ashley A. and {D'Avanzo}, Paolo and {D'Elia}, Valerio and {De Pasquale}, Massimiliano and {Dyer}, Martin J. and {Galloway}, Duncan K. and {Garbutt}, James A. and {Green}, Matthew J. and {Hartmann}, Dieter H. and {Jakobsson}, P{\'a}ll and {Kerry}, Paul and {Kouveliotou}, Chryssa and {Langeroodi}, Danial and {Le Floc'h}, Emeric and {Leung}, James K. and {Littlefair}, Stuart P. and {Munday}, James and {O'Brien}, Paul and {Parsons}, Steven G. and {Pelisoli}, Ingrid and {Sahman}, David I. and {Salvaterra}, Ruben and {Sbarufatti}, Boris and {Steeghs}, Danny and {Tagliaferri}, Gianpiero and {Th{\"o}ne}, Christina C. and {de Ugarte Postigo}, Antonio and {Kann}, David Alexander},
        title = "{Heavy-element production in a compact object merger observed by JWST}",
      journal = {\nat},
         year = 2024,
        month = feb,
       volume = {626},
       number = {8000},
        pages = {737-741},
          doi = {10.1038/s41586-023-06759-1},
archivePrefix = {arXiv},
       eprint = {2307.02098},
 primaryClass = {astro-ph.HE},
       adsurl = {https://ui.adsabs.harvard.edu/abs/2024Natur.626..737L}
}

@ARTICLE{Troja18b,
       author = {{Troja}, E. and {Piro}, L. and {Ryan}, G. and {van Eerten}, H. and {Ricci}, R. and {Wieringa}, M.~H. and {Lotti}, S. and {Sakamoto}, T. and {Cenko}, S.~B.},
        title = "{The outflow structure of GW170817 from late-time broad-band observations}",
      journal = {\mnras},
         year = 2018,
        month = jul,
       volume = {478},
       number = {1},
        pages = {L18-L23},
          doi = {10.1093/mnrasl/sly061},
archivePrefix = {arXiv},
       eprint = {1801.06516},
 primaryClass = {astro-ph.HE},
       adsurl = {https://ui.adsabs.harvard.edu/abs/2018MNRAS.478L..18T}
}

@ARTICLE{Fujibayashi20c,
       author = {{Fujibayashi}, Sho and {Shibata}, Masaru and {Wanajo}, Shinya and {Kiuchi}, Kenta and {Kyutoku}, Koutarou and {Sekiguchi}, Yuichiro},
        title = "{Viscous evolution of a massive disk surrounding stellar-mass black holes in full general relativity}",
      journal = {\prd},
         year = 2020,
        month = dec,
       volume = {102},
       number = {12},
          eid = {123014},
        pages = {123014},
          doi = {10.1103/PhysRevD.102.123014},
archivePrefix = {arXiv},
       eprint = {2009.03895},
 primaryClass = {astro-ph.HE},
       adsurl = {https://ui.adsabs.harvard.edu/abs/2020PhRvD.102l3014F}
}

@ARTICLE{Kouveliotou93,
       author = {{Kouveliotou}, Chryssa and {Meegan}, Charles A. and {Fishman}, Gerald J. and {Bhat}, Narayana P. and {Briggs}, Michael S. and {Koshut}, Thomas M. and {Paciesas}, William S. and {Pendleton}, Geoffrey N.},
        title = "{Identification of Two Classes of Gamma-Ray Bursts}",
      journal = {\apjl},
         year = 1993,
        month = aug,
       volume = {413},
        pages = {L101},
          doi = {10.1086/186969},
       adsurl = {https://ui.adsabs.harvard.edu/abs/1993ApJ...413L.101K}
}

@ARTICLE{Fujibayashi20b,
       author = {{Fujibayashi}, Sho and {Wanajo}, Shinya and {Kiuchi}, Kenta and {Kyutoku}, Koutarou and {Sekiguchi}, Yuichiro and {Shibata}, Masaru},
        title = "{Postmerger Mass Ejection of Low-mass Binary Neutron Stars}",
      journal = {\apj},
         year = 2020,
        month = oct,
       volume = {901},
       number = {2},
          eid = {122},
        pages = {122},
          doi = {10.3847/1538-4357/abafc2},
archivePrefix = {arXiv},
       eprint = {2007.00474},
 primaryClass = {astro-ph.HE},
       adsurl = {https://ui.adsabs.harvard.edu/abs/2020ApJ...901..122F}
}

@ARTICLE{Fujibayashi20a,
       author = {{Fujibayashi}, Sho and {Shibata}, Masaru and {Wanajo}, Shinya and {Kiuchi}, Kenta and {Kyutoku}, Koutarou and {Sekiguchi}, Yuichiro},
        title = "{Mass ejection from disks surrounding a low-mass black hole: Viscous neutrino-radiation hydrodynamics simulation in full general relativity}",
      journal = {\prd},
         year = 2020,
        month = apr,
       volume = {101},
       number = {8},
          eid = {083029},
        pages = {083029},
          doi = {10.1103/PhysRevD.101.083029},
archivePrefix = {arXiv},
       eprint = {2001.04467},
 primaryClass = {astro-ph.HE},
       adsurl = {https://ui.adsabs.harvard.edu/abs/2020PhRvD.101h3029F}
}

@ARTICLE{Shibata19,
       author = {{Shibata}, Masaru and {Hotokezaka}, Kenta},
        title = "{Merger and Mass Ejection of Neutron Star Binaries}",
      journal = {Annual Review of Nuclear and Particle Science},
         year = 2019,
        month = oct,
       volume = {69},
        pages = {41-64},
          doi = {10.1146/annurev-nucl-101918-023625},
archivePrefix = {arXiv},
       eprint = {1908.02350},
 primaryClass = {astro-ph.HE},
       adsurl = {https://ui.adsabs.harvard.edu/abs/2019ARNPS..69...41S}
}

@ARTICLE{Smaranikab22,
       author = {{Banerjee}, Smaranika and {Tanaka}, Masaomi and {Kato}, Daiji and {Gaigalas}, Gediminas and {Kawaguchi}, Kyohei and {Domoto}, Nanae},
        title = "{Opacity of the Highly Ionized Lanthanides and the Effect on the Early Kilonova}",
      journal = {\apj},
         year = 2022,
        month = aug,
       volume = {934},
       number = {2},
          eid = {117},
        pages = {117},
          doi = {10.3847/1538-4357/ac7565},
archivePrefix = {arXiv},
       eprint = {2204.06861},
 primaryClass = {astro-ph.HE},
       adsurl = {https://ui.adsabs.harvard.edu/abs/2022ApJ...934..117B}
}

@ARTICLE{Smaranikab24,    
       author = {{Banerjee}, Smaranika and {Tanaka}, Masaomi and {Kato}, Daiji and {Gaigalas}, Gediminas},
        title = "{Diversity of Early Kilonova with the Realistic Opacities of Highly Ionized Heavy Elements}",
      journal = {\apj},
         year = 2024,
        month = jun,
       volume = {968},
       number = {2},
          eid = {64},
        pages = {64},
          doi = {10.3847/1538-4357/ad4029},
archivePrefix = {arXiv},
       eprint = {2304.05810},
 primaryClass = {astro-ph.HE},
       adsurl = {https://ui.adsabs.harvard.edu/abs/2024ApJ...968...64B}
}

@ARTICLE{Lucy03,
       author = {{Lucy}, L.~B.},
        title = "{Monte Carlo transition probabilities. II.}",
      journal = {\aap},
         year = 2003,
        month = may,
       volume = {403},
        pages = {261-275},
          doi = {10.1051/0004-6361:20030357},
archivePrefix = {arXiv},
       eprint = {astro-ph/0303202},
 primaryClass = {astro-ph},
       adsurl = {https://ui.adsabs.harvard.edu/abs/2003A&A...403..261L}
}

@ARTICLE{Paczynski91,
       author = {{Paczynski}, Bohdan},
        title = "{Cosmological gamma-ray bursts.}",
      journal = {\actaa},
         year = 1991,
        month = jan,
       volume = {41},
        pages = {257-267},
       adsurl = {https://ui.adsabs.harvard.edu/abs/1991AcA....41..257P}
}
\bibliographystyle{aasjournal}
\end{document}